\documentclass[aps,prd,superscriptaddress,twocolumn,preprintnumbers]{revtex4-1}

\usepackage{SIunits}
\usepackage{amsmath}
\usepackage{subfig}
\usepackage{graphicx}
\usepackage{color}

\newcommand{\der}{\text{d}}
\newcommand{\virod}{\Delta_{\text{vir}}}
\newcommand{\delr}{\delta_\rho}
\newcommand{\delvs}{\delta_{v^2}}
\newcommand{\mean}[1]{\left\langle#1\right\rangle}
\newcommand{\fulljust}{\setlength{\rightskip}{0pt}\setlength{\leftskip}{0pt}}

\begin{document}

\title{Effects of Velocity-Dependent Dark Matter Annihilation on the
Energy Spectrum of the Extragalactic Gamma-ray Background}

\author{Sheldon Campbell}
\author{Bhaskar Dutta}
\affiliation{Department of Physics and Astronomy, Texas A\&M University, College Station, Texas 77843, USA}
\author{Eiichiro Komatsu}
\affiliation{Texas Cosmology Center and Department of Astronomy, The University of Texas at Austin, Austin, Texas 78712, USA}


\preprint{MIFPA-10-39}
\preprint{TCC-023-10}

\begin{abstract}
We calculate the effects of velocity-dependent dark matter annihilation cross sections on the intensity of the extragalactic gamma-ray background.  Our formalism does not assume a locally thermal distribution of dark matter particles in phase space, and is valid for arbitrary velocity-dependent annihilation.  As concrete examples, we calculate the effects of p-wave annihilation (with the $v$-weighted cross section of $\sigma v=a+bv^2$) on the mean intensity of extragalactic gamma rays produced in cosmological dark matter halos.  This velocity variation makes the \textit{shape} of the energy spectrum harder, but this change in the shape is too small to see unless $b/a\agt 10^6$.  While we find no such models in the parameter space of the Minimal Supersymmetric Standard Model (MSSM), we show that it is possible to find $b/a\agt 10^6$ in the extension MSSM$\otimes U(1)_{B-L}$.  However, we find that the most dominant effect of the p-wave annihilation is the suppression of the \textit{amplitude} of the gamma-ray background.  A non-zero $b$ at the dark matter freeze-out epoch requires a smaller value of $a$ in order for the relic density constraint to be satisfied, suppressing the amplitude by a factor as low as $10^{-6}$ for a thermal relic.  Non-thermal relics will have weaker amplitude suppression.  As another velocity-dependent effect, we calculate the spectrum for s-wave annihilation into fermions enhanced by the attractive Sommerfeld effect.  Resonances associated with this effect result in significantly enhanced intensities, with a slightly softer energy spectrum.
\end{abstract}
\maketitle

\section{Introduction}
The simplest cosmology that consistently explains astronomical observations to date is the $\Lambda$CDM scenario where about 84\% of the matter in the universe is the so-called dark matter \cite{Komatsu:2008hk,*Komatsu:2010fb}.  All current evidence for dark matter is based on its large-scale gravitational effects; details of its fundamental nature remain elusive and are the center of a major campaign in modern experimental physics.  Since the dark matter does not radiate,  it must be non-baryonic and electrically neutral.  Neutrinos can only account for a small fraction of the dark matter, owing to their small masses \cite{Lee:1977ua,*Sato:1977ye}.  Therefore, the current simplest explanation  is that the major component of dark matter is a new stable, fundamental, massive particle relic from the big bang that interacts weakly with the standard model particles.

Dark matter particles may be produced directly in collider experiments. They may also be detected directly in nuclear-recoil experiments. In addition, \textit{indirect} detection of dark matter particles from their decay products such as high-energy cosmic rays and gamma rays offers independent and promising probes of the nature of dark matter \cite{Bertone:2004pz}. Indeed, there are claims that excesses of the high-energy cosmic-ray flux over the expected astrophysical signals have been found: PAMELA reported an excess of positron fraction in the energy range of $\unit{\text{60--100}}{\giga\electronvolt}$ \cite{Adriani:2008zr}.  ATIC has seen a bump in the electron plus positron spectrum at about $\unit{500}{\giga\electronvolt}$ \cite{Chang:2008zzr} that has since been followed up and not seen by the calorimeter on the Fermi Gamma-ray Satellite Telescope \cite{Abdo:2009zk,*Ackermann:2010ij}.  The difficulty with interpreting these results in terms of annihilation or decay of dark matter particles is that there are possible astrophysical explanations for these observations.  For example, there could be additional emission of positrons from pulsars that are not currently included in standard estimations of local positron abundances \cite{Profumo:2008ms,*Grasso:2009ma}. There could also be positrons accelerated in supernova shocks around dying massive stars \cite{Biermann:2009qi}.

Another complication is the propagation of charged particles in the presence of interstellar magnetic fields and plasmas in the Milky Way. On the other hand, gamma rays do not suffer from this complication, and thus offer a relatively clean probe of possible dark matter annihilation/decay signatures in the universe. In this paper, we describe improvements to predictions of extragalactic gamma rays produced by dark matter annihilations in cosmological halos.  



To date, most of the calculations of the mean intensity and angular power spectrum of the extragalactic gamma-ray background have been done for s-wave dominated thermal relics with an example spectrum per annihilation motivated from the minimal supersymmetric standard model (MSSM) and the weak-scale annihilation cross section of $\sigma v=\unit{3\times10^{-26}}{\centi\cubic\metre\per\second}$ \cite{Ando:2005xg,*Ando:2006cr,*Cuoco:2007sh,*SiegalGaskins:2008ge}. (See \cite{Kistler:2009xf} for a study on the effect of the velocity-dependent Sommerfeld enhancement of the annihilation cross section in \textit{Galactic} subhalos.) New methods are needed to understand the effects that relative velocities between annihilating particles have on the observed gamma rays.  For example, there are many cases in the MSSM where momentum effects in the cross section must be properly implemented when calculating the relic density \cite{Gondolo:1990dk}.  

Instead of using the perturbation theory description of large-scale structure, it is more appropriate for this calculation to use the so-called ``halo approach'' \cite{Cooray:2002dia} because annihilations will occur predominantly in the densest regions of space, such as the cores of dark matter halos.  Within this framework, we develop a new method for modeling the distribution of one-point mean relative velocities of dark matter particles. Our method is based on universal halo properties observed in the simulations of large scale structure that have been upheld in recent simulations of individual halos.  

This paper is organized as follows. Our formalism is described in Section~\ref{sec:halo model}. As a first application of this formalism, we present a case study of p-wave annihilation in Section~\ref{sec:intensity}. In Section~\ref{sec:MSSM}, we investigate these effects within the context of the MSSM. In Section~\ref{sec:pdominated}, we also consider an example of a model for which the s-wave component of annihilation is negligible. Although adding only a p-wave term to the cross section is sufficient to describe the physics of many models, there are other interesting velocity-dependent scenarios such as Sommerfeld enhancements, Breit-Wigner resonances, and thresholds of new annihilation channels. It is conceivable that some of these effects will also change the spectrum per annihilation. In Section~\ref{sec:generalv}, we show how these effects can in general be incorporated into our framework. In Section~\ref{sec:sommerfeld}, we use our formalism to calculate the effects of Sommerfeld-enhanced s-wave annihilation into fermions on the extragalactic gamma-ray spectrum.  Our conclusions and the consequences of our findings are discussed in Section~\ref{sec:discussion}.

\section{The phase distribution of high density regions of dark matter}
\label{sec:halo model}

\subsection{Mass function}
The spherical halo model that we adopt in this paper approximates the distribution of matter as an ensemble of universal, disjoint, rigid, spherical halos.  This gives a good depiction of the densest regions of space, which are the overwhelmingly dominant sources of annihilation products.  Although halos are ellipsoidal in general, their cores are seen in simulations to be distributed nearly spherically.

The meaning of halo universality is that each halo's individual properties can be expressed in terms of universal halo variables that are assumed to fully specify the properties of the halo.  In general, halo variables may include quantities such as halo position, mass, cosmological redshift, concentration, formation redshift, internal angular momentum, etc.  Universal halo functions then depend only on the halo variables and position within the halo.  

The statistical description promotes halo variables to random variables with distributions from which ensemble averages of the halo functions can be determined.  Due to cosmological uniformity, statistical moments of halo properties are independent of halo position on a shell of fixed cosmological redshift.

The simplest and most maturely developed realization of this framework has only two halo variables: mass, $M$, and redshift, $z$, of halos.  The mean distribution of halos is well described by the Sheth-Tormen mass function $\frac{\der n}{\der M}(M,z)$ \cite{Sheth:1999su, *Sheth:2001dp}.  With this mass function, truncated halo profiles are used; hence, the rigid halo approximation: the density of matter not within any halo's virial radius $R_{\text{vir}}$ is neglected.

Following the analysis of \cite{White:2002at}, we use the virial mass-radius relation at redshift $z$ of
\begin{equation}
  \label{eq:MRrel}
  M=\frac{4}{3}\pi R_{\text{vir}}^3\virod\overline{\rho}(z)
\end{equation}
for the $z$-independent mean halo virial overdensity $\virod=180$. Here, $\overline{\rho}(z)$ is the background matter density at redshift $z$ (not the critical density). Note that the Sheth-Tormen mass function has been calibrated against N-body simulations with the mass defined by the Friends-of-Friends (FoF) algorithm. This introduces an ambiguity as to what value of $\virod$ corresponds to the FoF mass. However, we shall ignore this subtlety because this is not the dominant uncertainty in our calculation.

Although the halos in this model are not formally disjoint (even with the rigidity assumption), we neglect effects due to intersecting halos, assuming those regions to be rare in the model.  This allows us to ignore the negligible 2 and 3-halo contributions to the 1-point statistics calculated below in Section~\ref{sec:intensity}.

\subsection{Density profile}
The major breakthrough that makes the universal halo model viable is the discovery from large scale structure simulations that the spherically-averaged radial density of relaxed virialized halo structures follows a universal profile $\rho_h$ \cite{Navarro:1996gj}.  To facilitate easier comparison with previous results, we have decided in this paper to use the truncated NFW halo density profile \cite{Navarro:1996gj} 
\begin{equation}
  \label{eq:nfw}
  \rho_h(r|\rho_s,r_s,R_{\text{vir}})=
  \begin{cases}
    \frac{\rho_s}{\frac{r}{r_s}\left(1+\frac{r}{r_s}\right)^3}& \text{for $r<R_{\text{vir}}$},\\
    0& \text{otherwise.}
  \end{cases}
\end{equation}
Here, the halo profile is expressed in terms of the distance $r$ from the halo center, and the halo variables are the scale density $\rho_s$ and the scale radius $r_s$ of the halo.

It is worth noting, however, that recent simulations have had sufficient resolution and convergence to convincingly suggest that the NFW profile is too steep at $r\lesssim r_s$ \cite{Navarro:2008kc}.  Also, there are many new constraints on the concentration distribution from various simulations and astrophysical observations \cite{Duffy:2008pz}. The robustness of the calculation of the extragalactic signal to these astrophysical uncertainties will need to be explored in future work.

It is useful for what follows to explain how Eq.~(\ref{eq:nfw}) is expressed in terms of our halo variables $M$ and $z$.  An expression for $R_{\text{vir}}(M,z)$ is found from Eq.~(\ref{eq:MRrel}).  In the context of truncated NFW profiles, the halo concentration is defined to be $c\equiv R_{\text{vir}}/r_s$, from which we get
\begin{equation}
  \label{eq:rs}
  r_s(M,c,z)=\frac{1}{c}\left[\frac{3M}{4\pi\virod\overline{\rho}(z)}\right]^{\frac{1}{3}}.
\end{equation}
Finally, integrating over the density profile for the halo mass and solving for the scale density gives
\begin{equation}
  \label{eq:dens}
  \rho_s(c,z)=\frac{\virod\overline{\rho}(z)c^3}{3\left[\ln(1+c)-\frac{c}{1+c}\right]}.
\end{equation}
Taking each halo to be at the mean concentration $c=\overline{c}(M,z)$ from \cite{Bullock:1999he} gives us the expression for $\rho_h(r|M,z)$ used in the spherical halo model. 

While we shall adopt the mean concentration-mass relation of \cite{Bullock:1999he} throughout this paper, the next generalization of this model would be to determine a concentration-mass function $\frac{\der n}{\der M\der c}(M,c,z)$. Although it appears that halo concentrations are distributed via a log-normal distribution \cite{Bullock:1999he}, the full joint concentration-mass distribution needs to be determined.

\subsection{Velocity profile}
\label{sec:velocityprofile}
To insert a distribution of relative velocities in the spherical halo model, one would ideally want a universal spherically-averaged halo profile of particle velocities.  Such a profile was first observed in simulations by Taylor and Navarro, in 2001, in terms of the halo's density profile and its velocity variance profile $\sigma_{vh}^2(r|M,z)$.  They found that 
\begin{equation}
  \label{eq:phase density}
  \frac{\rho_h}{\sigma_{vh}^3}\propto r^{-\alpha}
\end{equation}
provides an excellent fit to their numerical simulation \cite{Taylor:2001bq}. Here, $\alpha$ is a constant. This result has recently been verified by the analysis of simulations such as the Aquarius simulations \cite{Navarro:2008kc} and GHALO \cite{Stadel:2008pn}, which have much better spatial resolution.

In general, the velocity dispersion at any position in the halo is not isotropic, but the simulations show that the anisotropy is small in the inner regions of halos, $r\lesssim r_s$. As the annihilation signal is dominated by these inner regions, we shall assume isotropic velocity dispersions. 

Since, in this case, the density and velocity dispersion of a collisionless self-gravitating system is described by the radial Jeans equation
\begin{equation}
  \label{eq:Jeans}
  \frac{\der}{\der r}\left[\frac{-r^2}{G\rho_h}\frac{\der(\rho_h\sigma_{vh}^2)}{\der r}\right]=4\pi\rho_hr^2.
\end{equation}
Eq.~(\ref{eq:phase density}) and Eq.~(\ref{eq:Jeans}) can be combined to solve for a family of density profiles and associated velocity dispersion profiles.  Taylor and Navarro found a critical value for the proportionality constant in Eq.~(\ref{eq:phase density}) that produced a density profile consistent with the universal profile seen in simulations.  Dehnen and McLaughlin \cite{Dehnen:2005cu} generalized this analysis, generating the Dehnen-McLaughlin profiles.  The ``critical profile'' (the only physical solution without an outer truncation where $\rho_h$ becomes negative) occurs at $\alpha=\frac{35}{18}$, which is consistent with the values measured from the Aquarius simulation \cite{Navarro:2008kc}.  The resulting profile is similar to an NFW profile, but is less steep at the halo core. Nevertheless, it is still consistent with the Aquarius simulation \cite{Navarro:2008kc}. 

To use this result for any other density profile consistent with simulations, we can simply treat the other profile as an approximation of the Dehnen-McLaughlin solution and use Eq.~(\ref{eq:phase density}) to solve for the associated halo velocity dispersion.  We now explain how to do this for the NFW profile.  This method can be applied to any other density profiles.

When matching different profiles, it is common practice  to match them at the halo radius $r_{-2}$ where the minus of the logarithmic slope of the density profile, $\gamma\equiv-\der\ln\rho_h / \der\ln r$, is $2$.  However, because the power law profile with $\gamma(r)=6-2\alpha$ is a solution to the equations and all other relevant solutions have a radius where $\gamma$ takes this value, it is more convenient to parametrize the Dehnen-McLaughlin profiles at the radius $r_0$ where
\begin{equation}
  \label{eq:r0}
  \gamma(r_0)\equiv6-2\alpha.
\end{equation}
Defining the halo variables $\rho_0\equiv\rho_h(r_0)$ and $\sigma_{v0}\equiv\sigma_{vh}(r_0)$, the family of Dehnen-McLaughlin profiles can then be parametrized by the dimensionless parameter
\begin{equation}
  \label{eq:kappa}
  \kappa\equiv\frac{4\pi G\rho_0r_0^2}{\sigma_{v0}^2},
\end{equation}
which fixes the proportionality constant in Eq.~(\ref{eq:phase density}).  The critical NFW-like profile for $\alpha=\frac{35}{18}$ occurs for $\kappa=\frac{200}{81}$. 

We match the density profiles at $r_0$ by differentiating the NFW profile [Eq.~(\ref{eq:nfw})]
\begin{equation}
  \gamma(r)=\frac{2r}{r+r_s}+1
\end{equation}
and use Eq.~(\ref{eq:r0}) to find
\begin{equation}
  r_0(M,c,z)=\frac{5-2\alpha}{2\alpha-3}r_s(M,c,z).
\end{equation}
Putting this back into $\rho_h(r_0)$, we find
\begin{equation}
  \rho_0(c,z)=\frac{(2\alpha-3)^3}{4(5-2\alpha)}\rho_s(c,z).
\end{equation}
Finally, we use Eq.~(\ref{eq:phase density}) and Eq.~(\ref{eq:kappa}) to get the desired universal halo velocity variance profile as
\begin{eqnarray}
\nonumber
  \sigma_{vh}^2(r|M,c,z)&=&\frac{4\pi
   G}{\kappa}\rho_0^{\frac{1}{3}}(c,z)r_0^2(M,c,z) \\
&\times& \left[\left(\frac{r}{r_0(M,c,z)}\right)^\alpha\rho_h(r|M,c,z)\right]^{\frac{2}{3}}\!\!\!\!\!.
\end{eqnarray}

\subsection{Derived relative velocity and mean annihilation cross-section profiles}
\label{sec:relv}

The dependence of the annihilation cross section on the center of mass energy of the annihilating particles can be written so that $\sigma v$ is a function of the square relative velocity $v^2$.  For dark matter particles that have a p-wave component of annihilation, the velocity-weighted annihilation cross section is given by
\begin{equation}
  \label{eq:cross section}
  \sigma v=a+bv^2
\end{equation}
where $a$ and $b$ are constant coefficients and $v$ is the relative
velocity between the annihilated particles\footnote{More precisely, $v$
is the relative velocity in the center of mass frame, but relativistic
corrections will be negligible.  See \cite{Gondolo:1990dk}, for
example.}.  

Note that, in the theory of relativistic partial wave analysis, the s-wave component of expansion contributes to both $a$ and $b$ \cite{Drees:1992am}. However, it now appears to be common in the literature to simply refer to $a$ as the s-wave component and $b$ as the p-wave component. This is fine when the p-wave contribution to $b$ is large compared to that of the s-wave, and this will be the meaning of those terms in the context of this paper.  One just needs to keep in mind that, when the p-wave component of the partial wave is small, $b$ will technically be dominated by the s-wave component of the partial wave.  Nevertheless, the effect of $b$ from the s-wave contribution is negligible in relic density and gamma-ray signal calculations, since it is velocity suppressed.

Since the relevant particle physics depends on the square relative velocity between annihilating particles, we are interested in a spherically-averaged halo profile of the mean squared relative velocity $v_h^2(r)$ at each radial position within the halo.  This depends on the underlying distribution $f_{\mathbf{u}}(\mathbf{u}_1,\mathbf{r})$ of particle velocities $\mathbf{u}_1$ at position $\mathbf{r}$.  The probability distribution of square relative velocities at that position can be calculated from
\begin{equation}
  f_{v^2}(v_1^2,\mathbf{r})=\int\der^3\mathbf{u}_1\der^3\mathbf{u}_2f_{\mathbf{u}}(\mathbf{u}_1,\mathbf{r})f_{\mathbf{u}}(\mathbf{u}_2,\mathbf{r})\delta(v_1^2-|\mathbf{u}_1-\mathbf{u}_2|^2),
\end{equation}
where $\delta$ is the Dirac delta function, and the mean square relative velocity at that position is
\begin{equation}
  v^2(\mathbf{r})=\int\der(v_1^2) v_1^2 f_{v^2}(v_1^2,\mathbf{r}).
\end{equation}
When the underlying velocity distribution $f_\mathbf{u}$ is thermal with variance $\sigma_{vh}^2(r)$ at each halo position $r$, we find $v_h^2(r)=6\sigma_{vh}^2(r)$.  But halo simulations show significant deviations from Maxwell-Boltzmann distributions \cite{Vogelsberger:2008qb,*Zemp:2008gw}.  Therefore, we will suppose that the universal phase distribution satisfies 
\begin{equation}
  v_h^2(r)=\lambda\sigma_{vh}^2(r)
\end{equation}
for some constant $\lambda$. The applicability of this relation should be verified by the current simulation data, which would give the value of $\lambda$. In this paper, we shall leave this as a free parameter, and use $\lambda=6$ in those calculations that require an explicit value.

The mean relative-velocity-weighted cross section at each position is no longer a thermal average in general, and can even be taken as a universal halo function.  The mean weighted cross section at position $r$ of a halo is
\begin{equation}
  \label{eq:halo cs}
  [\sigma v]_h(r)=\int\der(v_1^2)\;[\sigma v](v_1^2)\;f_{v^2}(v_1^2,r)
\end{equation}
which, in our particular case study of Eq.~(\ref{eq:cross section}), is given by
\begin{equation}
  [\sigma v]_h(r)=a+\lambda b\sigma_{vh}^2(r).
\end{equation}

\subsection{Further comments on modeling the dark matter distribution}
\label{sec:dist details}

For the sample calculations in this paper, we have assumed $\Lambda$CDM cosmological parameters from WMAP5 \cite{Komatsu:2008hk} neglecting neutrino effects: $\Omega_\Lambda=0.721$, $\Omega_b=0.0462$, $\Omega_c=1-\Omega_\Lambda-\Omega_b$, $h=0.701$, $\sigma_8=0.817$, and $n_s=0.96$.  When calculating the halo mass function, we used the linear power spectrum proposed by Eisenstein and Hu \cite{Eisenstein:1997jh}, and a critical overdensity for spherical collapse of 1.686.

If the Milky Way Galaxy is in an average halo of mass $\unit{2.0\times10^{12}}{M_\odot}$ and the Solar System is $\unit{8.0}{\kilo pc}$ from the center, then this model gives the local dark matter particle density to be $\unit{4.1\times10^{-25}}{\gram\per\centi\meter\cubed}=\unit{0.23}{\giga\electronvolt\per\centi\meter\cubed}$, and the rms relative particle velocity to be $\unit{0.85\times10^{-3}\sqrt{\lambda/6}}{c}$. For comparison, the latest determination of the local dark matter density in our solar neighborhood, based on data from both dynamical observations and galactic simulations, gave $\unit{0.47\pm 0.03\pm 0.08}{\giga\electronvolt\per\centi\meter\cubed}$ \cite{Pato:2010yq} (the first uncertainty is statistical while the second one is systematic).  This value takes into account the systematic effect of the dark disk structure which the spherical model neglects, so some underestimation by our model is expected.  Thus, our model is able to reproduce the observed value reasonably well.

We have made a number of additional simplifications while calculating the example spectra in this paper:

\begin{itemize}
\item [1.]
The results in our model are sensitive to the minimum mass of halos $M_{\text{min}}$, since the cores of NFW halos become more dense as the halo mass decreases.  This low mass halo scale is determined by the Jeans mass, which can be determined for any given particle physics model \cite{Bertschinger:2006nq, *Bringmann:2009vf, *Kasahara:2009th}.  These works show that $10^{-12} M_\odot \alt M_{\text{min}} \alt 10^{-3} M_\odot$ in minimal supersymmetric extensions of the standard model.  For the calculations in this paper, we set $M_{\text{min}}=10^6 M_\odot$.  Calculations using a smaller minimum mass increased the intensity by a factor less than 10 (see Appendix~\ref{ap:smallminmass}).

\item[2.]
This initial work also neglects the contributions of halo substructures.  Recent simulations have resolved the abundance of halo substructures down to $10^5 M_\odot$ \cite{Diemand:2006ik, *Diemand:2008in, *Springel:2008cc}.  These results are easily implemented in our models with recent new analytic methods for calculating the boost in intensity of annihilation gamma rays \cite{Martinez:2009jh, *Afshordi:2009hn}.  This work is in progress.  The halo substructure also suffers uncertainties due to the magnitude of $M_{\text{min}}$, since it also sets the minimum mass scale of sub-halos.

\item[3.]
The peculiar motions of halos, and the velocity of the annihilation center of mass relative to the halo have been neglected in this work.  The corresponding corrections to our spectra due to these redshifting effects will be small.

\item[4.]
While we have included the opacity of the universe to gamma rays in our expressions below, we will neglect the effect of opacity in the calculations presented in this paper.  Our ignoring the opacity does not change any of the conclusions of this paper: the suppression of photon intensity is quite small for photons below $\unit{10}{\giga\electronvolt}$; therefore, its effects will only be seen for large dark matter masses.  Using the fitting formula of Steckel, et al. \cite{Stecker:2005qs, *Stecker:2006eh} for the opacity, we checked its effect for a sample model having neutralino dark matter with a mass of $\unit{550}{\giga\electronvolt}$, which is the largest dark matter mass appearing in the sample calculations of this paper.  The extragalactic annihilation photon intensity spectrum peaked at about $\unit{20}{\giga\electronvolt}$ and the opacity began reducing the intensity more than 20\% only at photon energies above $\unit{40}{\giga\electronvolt}$, and 50\% reduction at energies above $\unit{200}{\giga\electronvolt}$.  For models with smaller dark matter masses, for instance at $\unit{150}{\giga\electronvolt}$, the opacity effect on the gamma-ray spectrum is especially small.  
\end{itemize}

\section{Extragalactic gamma-ray background: p-wave annihilation}
\label{sec:intensity}

\subsection{Formalism}

The specific intensity of gamma rays of energy $E_\gamma$ from p-wave annihilation of dark matter particles in the direction of $\hat{n}$ is given by \cite{Ando:2005xg} 
\begin{eqnarray}
\nonumber
& &  I_\gamma(\hat{n},E_\gamma)\\
\nonumber
&\simeq&\int\der r\ [\delr(\hat{n},r)+1]^2\ \frac{[\sigma
v](\hat{n},r)}{[\sigma v]_0}\ W\!\left((1+z)E_\gamma,z\right) \\
\nonumber
  &\simeq&\int\frac{\der z}{H(z)}\ \delta_\rho^2(\hat{n},z)\left[1+\frac{b}{a}\ v^2(\hat{n},z)\right]W\!\left((1+z)E_\gamma,z\right)\\
\end{eqnarray}
where
\begin{equation}
  \delr(\mathbf{r})\equiv\frac{\rho(\mathbf{r})}{\overline{\rho}(z)}-1\simeq\frac{\rho(\mathbf{r})}{\overline{\rho}(z)}
\end{equation}
is the dark matter overdensity.  In all regions that contribute non-negligibly to the intensity, it is true that $\rho(\mathbf{r})$, the dark matter density at position $\mathbf{r}$, is much larger than $\overline{\rho}(z)$, the background density at redshift z associated with the radial distance $|\mathbf{r}|$ from the observer. Here, $[\sigma v](\mathbf{r})=a+bv^2(\mathbf{r})$ is the mean relative-velocity-weighted dark matter annihilation cross section at position $\mathbf{r}$ \footnote{We do not use the angle bracket notation with the mean one-point velocity-weighted cross section to prevent confusion, since that notation is traditionally used to denote a thermal average in the literature. Also, the angle brackets will predominantly be used in this paper to denote averages over ensembles of dark matter halos.}; $[\sigma v]_0$ is a reference cross section that, in this context, is taken to be what we are referring to as the s-wave component of the cross section, i.e., $[\sigma v]_0\equiv\lim_{v\rightarrow 0}\sigma v=a$; $v^2(\mathbf{r})$ is the mean relative squared velocity of the dark matter at $\mathbf{r}$; and $H(z)$ is the Hubble function.  The intensity window function $W$ is given by
\begin{equation}
  \label{eq:window}
  W(E_\gamma,z)=\frac{1}{8\pi}\,[\sigma v]_0\,n_{\text{DM}}^2\,(1+z)^3\,\frac{\der N_\gamma(E_\gamma)}{\der E_\gamma}\,e^{-\tau(E_\gamma,z)}
\end{equation}
where $n_{\text{DM}}$ is the background dark matter number density today, $\frac{dN_\gamma}{dE_\gamma}$ is the photon spectrum per annihilation, and $\tau$ is the opacity of the universe to gamma rays.

The mean intensity profile is determined from joint statistical moments of the density and relative velocity fields.  In our quest to quantify the new velocity-dependent effects, we decompose the mean intensity into velocity-independent and velocity-dependent terms:
\begin{equation}
  \mean{I_\gamma}(E_\gamma)=\mean{I_\gamma}_0(E_\gamma)+\mean{I_\gamma}_v(E_\gamma)
\end{equation}
where
\begin{equation}
  \label{eq:S-Intensity}
  \mean{I_\gamma}_0(E_\gamma)=\int\frac{\der z}{H(z)}\mean{\delr^2}(z)\,W((1+z)E_\gamma,z)
\end{equation}
is the s-wave approximation of the mean intensity (valid when $b\ll a$) and
\begin{equation}
  \mean{I_\gamma}_v(E_\gamma)=\frac{b}{a}\int\frac{\der z}{H(z)}\mean{\delr^2v^2}(z)\,W((1+z)E_\gamma,z),
\end{equation}
which we refer to as the p-wave component.  

Note that had we not approximated $\delr+1\simeq\delr$, there would be an additional term with $\mean{\delr v^2}$, but this is always negligible when compared to the term with $\mean{\delr^2 v^2}$.  These averages can be calculated in the context of the spherical rigid halo model by taking an ensemble average over the positions and masses of halos that is consistent with the distributions seen in the cosmological simulations.  

\begin{figure*}[t]
\subfloat[]{\label{subfig:velvariance}\includegraphics[width=0.45\textwidth]{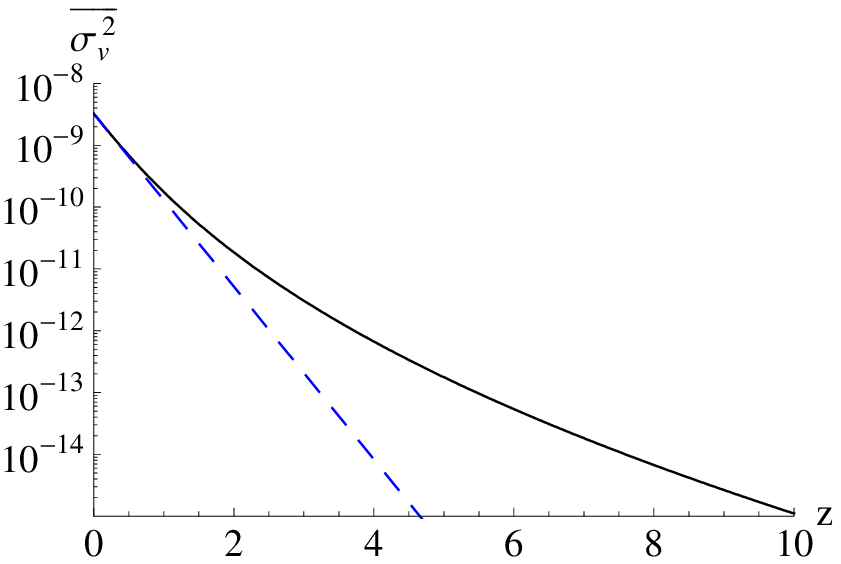}}
  \subfloat[]{\label{subfig:velvarapp}\includegraphics[width=0.45\textwidth]{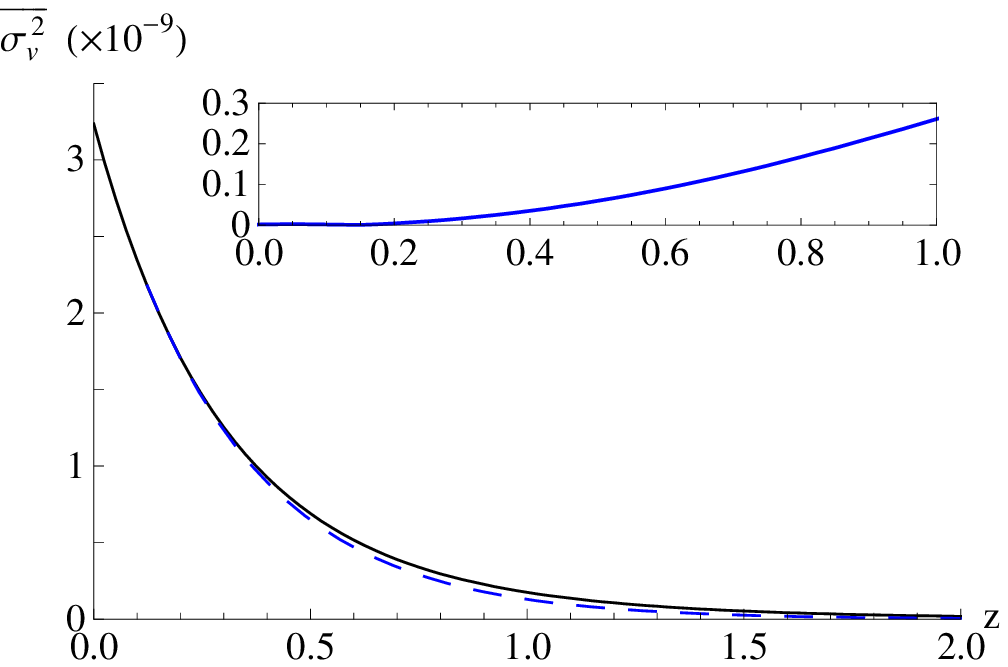}}
  \caption{\label{fig:mean square velocity}\fulljust 
Spatially-averaged one-point velocity variance, $\overline{\sigma_v^2}$, calculated from the spherical halo model (see Section~\ref{sec:velocityprofile}) for the cosmology described in Section~\ref{sec:dist details}. At low redshift, $z\ll1$, the variance is exponential in $z$, as indicated by the dashed curve given by $\overline{\sigma_v^2}(z)\simeq\overline{\sigma_v^2}(0)B^z$ with $B=0.04$.  The inset in the right plot shows the magnitude of the relative error of the exponential curve.\hfill\mbox{}}
\end{figure*}

The statistical moments are more convenient to evaluate for random variables that have vanishing mean.  To facilitate this, we define the mean-square-relative-velocity overdensity
\begin{equation}
  \delvs(\hat{n},z)\equiv\frac{v^2(\hat{n},z)}{\overline{v^2}(z)}-1,
\end{equation}
where $\overline{v^2}(z)$ is the average square one-point relative velocity at redshift $z$, calculated in the spherical halo model via
\begin{equation}
  \overline{v^2}(z)=\int\der^3\mathbf{r}\der M\frac{\der n}{\der M}(M,z)v_h^2(r|M,z),
\end{equation}
or in terms of the mean velocity variance at redshift $z$,
\begin{eqnarray}
\nonumber
  \overline{\sigma_v^2}(z)&=&\frac{\overline{v^2}(z)}{\lambda}\\
 &=&4\pi\int\der M\frac{\der n}{\der M}(M,z)\int\der r~r^2\sigma_{vh}^2(r|M,z).
\end{eqnarray}
A plot of $\overline{\sigma_v^2}(z)$ calculated with this model is shown by the solid curve in Figure~\ref{fig:mean square velocity}.  This curve is not sensitive to the uncertainty in $M_\text{min}$, since small mass halos contribute very little to the velocity variance.  Decreasing $M_\text{min}$ from $10^6 M_\odot$ to $10^{-6} M_\odot$ increases $\overline{\sigma_v^2}(0)$ by $2\times10^{-5}\%$ and even $\overline{\sigma_v^2}(10)$ by only 1\%.  

The fact that we see relative velocities increasing with time accentuates the fact that the particle motions are virial rather than thermal---increasing velocity variance is an indication of the growth of halo structures.  In the model we used, the rms relative velocity today (due to matter in high density regions) is calculated to be $\sqrt{\overline{v^2}(0)}=1.4\times10^{-4}\sqrt{\lambda/6}$.  At low redshift, we observe that the velocity variance is approximately exponential in $z$, given by $\overline{\sigma_v^2}(z)\simeq\overline{\sigma_v^2}(0)B^z$ for $B\approx0.04$.  This is indicated by the dashed curve in Figure~\ref{fig:mean square velocity} and shown to be accurate within 30\% up to $z=1$.  Note that all of the calculations in this paper use the full model for $\overline{\sigma_v^2}(z)$, and not the exponential approximation.

In terms of only overdensity random variables, the mean intensity is written as
\begin{eqnarray}
\nonumber
  \mean{I_\gamma}(E_\gamma)&=&\int\frac{\der z}{H(z)}\left[
    \mean{\delr^2}(z)+\beta(z)\mean{\delr^2\delvs}(z)\right]\\
&\times& W_v((1+z)E_\gamma,z)
  \label{intensity}
\end{eqnarray}
where we generalized the intensity window function
\begin{equation}
  W_v(E_\gamma,z)\equiv\left(1+\frac{b}{a}\lambda\overline{\sigma_v^2}(z)\right)W(E_\gamma,z),
\end{equation}
and introduced a new velocity coupling
\begin{equation}
  \beta(z)\equiv\left[1+\left(\frac{b}{a}\lambda\overline{\sigma_v^2}(z)\right)^{-1}\right]^{-1}
\end{equation}
that vanishes as $b\rightarrow0$ and is $1$ as $a\rightarrow0$.

After ignoring effects due to regions of intersecting halos, the
statistical moments we need are simply given by
\begin{align}
  \mean{\delr^2}(z)=&\int\!\der^3\mathbf{r}\der M\frac{\der n}{\der M}(M,z)\frac{\rho_h^2(r|M,z)}{\overline{\rho}^2(z)},\\ 
  \mean{\delr^2\delvs}(z)=&\int\!\der ^3\mathbf{r}\der M\frac{\der n}{\der M}(M,z)\frac{\rho_h^2(r|M,z)\sigma_{vh}^2(r|M,z)}{\overline{\rho}^2(z)\overline{\sigma_{v}^2}(z)},
\end{align}
where the mass integration is taken from our minimum halo mass $M_{\textrm{min}}=10^6 M_\odot$ to the maximum halo mass allowed by the halo concentration distribution \footnote{In the Bullock et al. distribution of halo concentrations, there is a halo mass above which the mean halo concentration becomes negative.} (determined to be $2\times10^{15} M_\odot$ for the calculations in this paper).

This mean intensity calculation is easily generalized for any relative velocity dependence of the particle physics model, although current derivations of the formalism require the velocity dependence to be at least piecewise analytic.  This will be discussed in Section~\ref{sec:generalv}.

\subsection{The contribution of the p-wave component to the mean annihilation gamma-ray intensity for arbitrary $\mathbf{\frac{b}{a}}$}
\label{sec:inten2ndterm}

The effect of the p-wave component of the cross section on the mean intensity of the gamma rays in this model is an energy-dependent intensity boost, relative to the s-wave approximation in Eq.~(\ref{eq:S-Intensity}), given by:
\begin{equation}
  \label{eq:p-wave boost}
  \frac{\mean{I_\gamma}(E_\gamma)}{\mean{I_\gamma}_0(E_\gamma)} = 1+\lambda\frac{b}{a}\Delta_I(E_\gamma)
\end{equation}
where we define
\begin{equation}
  \label{eq:DeltaI}
  \Delta_I(E_\gamma) \equiv \frac{\int\!\frac{\der z}{H(z)}\overline{\sigma_v^2}(z)\mean{\delr^2(1+\delvs)}\!(z)\,W((1+z)E_\gamma,z)}{\int\!\frac{\der z}{H(z)}\mean{\delr^2}\!(z)\,W((1+z)E_\gamma,z)}.
\end{equation}
Note that this depends only on the photon spectrum per annihilation, the dark matter phase space distribution, and the opacity.

Given that typical virial speeds for dark matter particles at the present epoch are somewhere around $\overline{u_0}\sim10^{-3} c$, one would only expect p-wave effects in the cross section to become important when $\frac{b}{a}\agt10^6$.  Applying this reasoning to Eq.~(\ref{eq:p-wave boost}), one would expect 
\begin{equation}
  \label{eq:approx deltaI}
  \lambda\Delta_I\sim\overline{u_0}^2\sim10^{-6}\Longrightarrow\Delta_I\sim10^{-7}.
\end{equation}

\begin{figure}[t]
  \subfloat[]{\label{subfig:DeltaI}\includegraphics[width=0.45\textwidth]{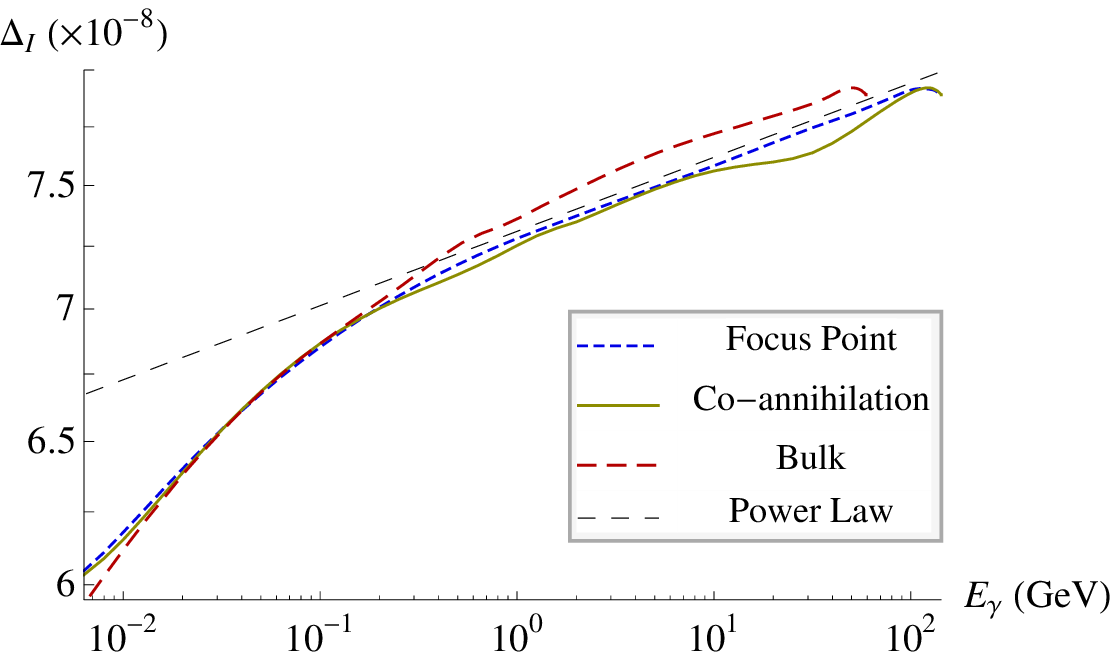}} \\
  \subfloat[]{\label{subfig:Inten}\includegraphics[width=0.45\textwidth]{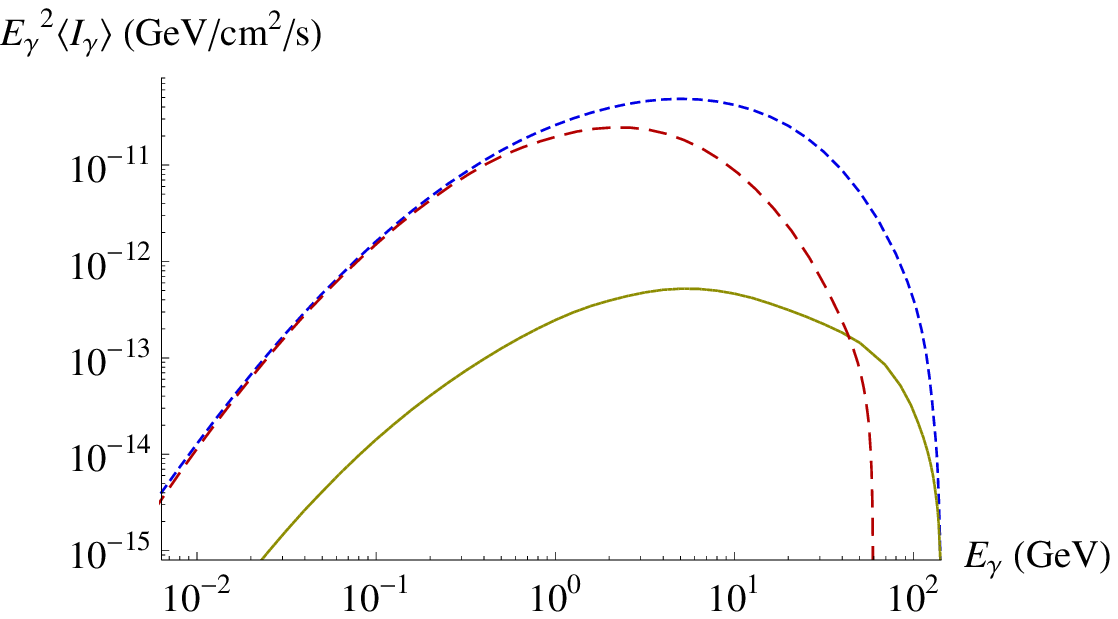}}
  \caption{\label{fig:tanb10example}\fulljust Top: $\Delta_I$ vs. $E_\gamma$ for a series of three mSUGRA models (at $\tan\beta=10$, $A_0=0$, and $\mu>0$) having neutralino dark matter with properties given in Table~\ref{tab:msugra}.  For comparison, a power law $\propto E_\gamma^{0.018}$ is also plotted.  Bottom: $E_\gamma^2\mean{I_\gamma}$ vs. $E_\gamma$ for those three models.\hfill\mbox{}}
\end{figure}

\begin{table*}
  \begin{ruledtabular}
  \begin{tabular}{ccccccccc}
    \ mSUGRA Region\ \ & $m_0$ (GeV) & $m_{1/2}$ (GeV) & $m_{\tilde{\chi}_1^0}$ (GeV) & $a$ ($\unit{\times10^{-26}}{\centi\meter\cubed\per\second}$) & $\frac{b}{a}$ & $\ \Omega_{c}h^2\ $ & $\ \ x_f\ \ $ & $\overline{\sigma v}(z_f)$ ($\unit{\times10^{-26}}{\centi\meter\cubed\per\second}$)\\
    \hline 
    Focus Point & 2569 & 395 & 150 & 1.9 & 1.8 & 0.114 & 22.9 & 2.6\\
    Bulk & 79 & 171 & 62.3 & 0.27 & 57.5 & 0.114 & 22.5 & 3.8\\
    Co-annihilation & 79 & 373.7 & 150 & 0.0019 & 378.8 & 0.113 & 24.0 & 5.8\\
  \end{tabular}
  \end{ruledtabular}
  \caption{\label{tab:msugra}\fulljust Sample mSUGRA models with parameters $\tan\beta=10$, $A_0=0$, and $\mu>0$.  The first two columns show the input model parameters.  The other columns show the calculated dark matter properties that are relevant to our discussion.  The last column shows the thermally-averaged $\sigma v$ at freezeout, including co-annihilations.\hfill\mbox{}}
\end{table*}

Figure~\ref{subfig:DeltaI} plots $\Delta_I(E_\gamma)$ for some sample
minimal supergravity (mSUGRA) models, also known as the constrained
minimal supersymmetric standard model (CMSSM).  In this paper, all MSSM
particle spectra, cross sections, annihilation spectra, and relic
densities are calculated using DarkSUSY~5.0.5 \cite{Gondolo:2004sc},
interfaced with ISAJET~7.78 \cite{Paige:2003mg}, and FeynHiggs~2.6.5.1
\cite{Hahn:2009zz}.  The values of $\Delta_I$ were determined
numerically to within a precision of 0.01\%.  The plot verifies the
expected order of magnitude for $\Delta_I$, and we see it is only mildly
energy-dependent, slowly increasing monotonically.  Over the energies
where the intensity spectrum peaks, $\Delta_I$ approximately follows a
power law relationship, of logarithmic slope 0.018, with the photon
energy.  The variations in minor structures between the curves are due
to differences in the particle spectrum per annihilation.

After satisfying all experimental constraints, most of the allowed parameter space is in four regions: 
\begin{itemize}
\item[1.] the focus point/hyperbolic region,
with larger higgsino component in the lightest neutralino
\cite{Chan:1997bi,*Feng:1999mn,*Feng:1999zg,*[{see also
}]Baer:1995nq,*Baer:1995va,*Baer:1998sz}; 
\item[2.] the stau-neutralino
co-annihilation region
\cite{Ellis:1998kh,*Ellis:1999mm,*Gomez:1999dk,*Gomez:2000sj,*Lahanas:1999uy,*Arnowitt:2001yh};
\item[3.] the Bulk region, where universal scalar and gaugino masses are small;
and 
\item[4.] the heavy Higgs/A annihilation funnel, where the lightest neutralino
mass is approximately twice the pseudoscalar Higgs mass
\cite{Drees:1992am,Baer:1997ai,*Baer:2000jj,*Ellis:2001msa,*Roszkowski:2001sb,*Djouadi:2001yk,*Lahanas:2001yr}.
\end{itemize}
Among these four regions, we have given examples of the first three in the parameter space where $\tan\beta=10$, $A_0=0$, and $\mu>0$. Properties of those specific models are listed in Table~\ref{tab:msugra}.  Most of the Bulk region parameter space is ruled out by current experimental constraints; the Bulk region example we are presenting is on the edge of those constraints.  

Figure~\ref{subfig:Inten} shows the annihilation gamma-ray spectra.  Since $\lambda b/a$ is much smaller than $\Delta_I^{-1}$ for each of these models (see the 6th column of Table~\ref{tab:msugra}), the velocity term of the intensity contributes negligibly to these spectra.  However, the p-wave strength $b/a$ of the cross section does have consequences for the normalization of the intensity curve, especially for the co-annihilation region. We shall explain this in the next subsection and in Section~\ref{sec:MSSM}. 

\subsection{P-wave suppression effect on the amplitude of the mean annihilation gamma-ray intensity}
\label{sec:intenrd}

The relic density of dark matter has reached the status of being an important constraint of particle physics models, owing to the fact that the magnitude of the relic density in our universe is precisely determined \cite{Komatsu:2008hk,*Komatsu:2010fb}, and that the technology needed to calculate the relic density for any particle physics model is mature. 

In the scenario of a thermally produced dark matter relic, numerical solutions of the Boltzmann equation present an approximate picture that holds, up to numerical corrections, typically within an order of magnitude: dark matter particles interact in thermal equilibrium with the big bang plasma until the thermal temperature is too low to produce new dark matter particles and the number density of dark matter particles is low enough so that annihilations are rare \cite{Lee:1977ua,*Sato:1977ye}.  After this time of freezeout, the number of dark matter particles in the universe per comoving volume is essentially constant (except for some residual annihilation occurring at very low annihilation rates).  

Therefore, the relic density magnitude can be accounted for, except for some important exceptions, by two quantities: the freezeout temperature normally expressed as $x_f=m_{\text{DM}}/T_f$, and the mean value of the dark matter annihilation cross section at freezeout $[\sigma v]_f$.  To thermally produce the correct relic density, the approximate value of the velocity weighed annihilation cross section needs to be $[\sigma v]_f\approx\unit{3\times10^{-26}}{\centi\meter\cubed\per\second}$.  

One important exception to the above argument is the following: if co-annihilations are present, then the annihilation cross section at freezeout is replaced by the larger effective co-annihilation cross section and the resulting relic density is lower.  For precision, we take the definition of $x_f$ to be at the thermal temperature where the dark matter number density is twice its thermal equilibrium value.

When in thermal equilibrium, the mean square relative velocity of the dark matter is related to the thermal temperature by $\overline{v^2}=6/x_f$.  Then, insofar as the approximate picture is valid, to meet the dark matter relic density constraint, a theory with p-wave annihilation strength $b/a$ must have an s-wave component of approximately
\begin{equation}
  a\approx\frac{[\sigma v]_f}{1+\left(\frac{b}{a}\right)\frac{6}{x_f}}
\end{equation}
in the absence of co-annihilations.  That is, the presence of p-wave annihilation requires the theory to have a smaller s-wave component in order to satisfy the relic density constraint.  This is the p-wave suppression effect.

If we were to compare the mean intensity of extragalactic annihilation photons with energy $E_\gamma$ in an s-wave model with $b/a=0$ to another model with the same annihilation spectrum, dark matter particle mass, and freezeout temperature, but with p-wave strength $b/a$, then the intensity due to a thermal relic would need to be p-wave suppressed by an approximate factor of
\begin{equation}
  \label{eq:pwavesupp}
  \frac{\mean{I_\gamma}\left(E_\gamma|\frac{b}{a}\right)}{\mean{I_\gamma}\left(E_\gamma|0\right)}\approx\frac{1+\left(\frac{b}{a}\right)\lambda\Delta_I(E_\gamma)}{1+\left(\frac{b}{a}\right)\frac{6}{x_f}}
\end{equation}
to satisfy the relic density constraint.  Here, when calculating $\Delta_I$ in this expression, we use the cross section at freezeout, $[\sigma v]_f$, 
as the reference cross section $[\sigma v]_0$ appearing in the window function
given in Eq.~(\ref{eq:window}).

\begin{figure}[t]
  \subfloat[]{\label{subfig:pwavesupp}\includegraphics[width=0.45\textwidth]{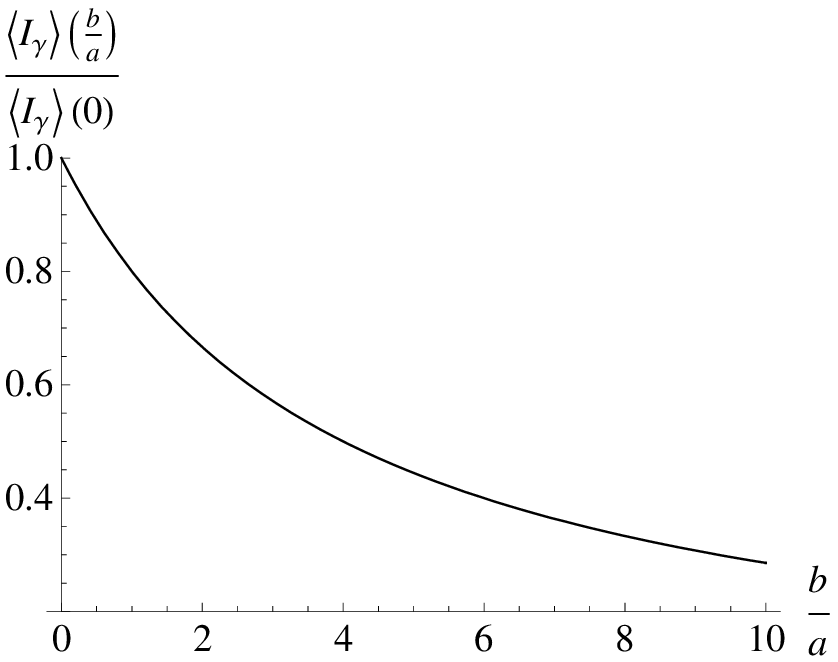}} \\
  \subfloat[]{\label{subfig:pwavesupplog}\includegraphics[width=0.45\textwidth]{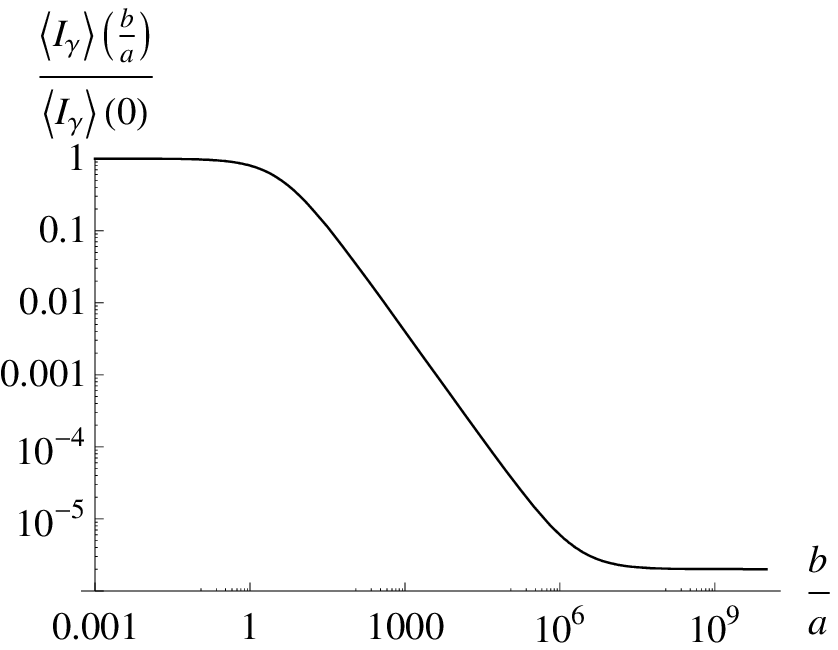}}
  \caption{\fulljust Top: The p-wave suppression factor for the mean intensity of extragalactic dark matter annihilation photons [Eq.~(\ref{eq:pwavesupp})] as a function of $b/a$, for typical values of $x_f=24$ and $\lambda\Delta_I=5\times 10^{-7}$.  The suppression is shown here for small values of $b/a$.  Bottom: The same plot, but on log scale and for a wider range of $b/a$.\hfill\mbox{}} 
\end{figure}

This effect is plotted for mild values of $b/a$ in Figure~\ref{subfig:pwavesupp} for $x_f=24$ and $\lambda\Delta_I=5\times10^{-7}$.  Since $\lambda\Delta_I\ll 6/x_f$, there are 3 regions of interest as can be seen explicitly in Figure~\ref{subfig:pwavesupplog}:
\begin{itemize}
\item[1.] When $b/a\ll x_f/6 \sim 4$, the theory is s-wave dominated, and the p-wave contributes very little to both the relic density calculation and the annihilation photon intensity spectrum.  We find when $b/a \alt1$, the intensity suppression is linear in $b/a$ with slope $-6/x_f\sim-1/4$.  
\item[2.] When $x_f/6\ll b/a\ll(\lambda\Delta_I)^{-1}$, it is important to include the p-wave for the accurate calculation of the relic density, resulting in a suppression of the s-wave of the annihilation cross section.  But the p-wave contribution to the shape of the intensity spectrum is negligible in this region.   
\item[3.] When $b/a\gg(\lambda\Delta_I)^{-1}\sim10^6$, the theory is p-wave dominated and the s-wave component contributes little to both calculations.  In this regime, p-wave suppression is maximal with a suppression factor of $x_f\lambda\Delta_I/6\sim x_f\Delta_I\sim10^{-6}$ relative to an equivalent s-wave dominated theory.  
\end{itemize}
In summary, for each of the three regions, the p-wave suppression goes like
\begin{equation}
  \frac{\mean{I_\gamma}\left(E_\gamma,\frac{b}{a}\right)}{\mean{I_\gamma}\left(E_\gamma,0\right)}\approx
  \begin{cases}
    1-\frac{6}{x_f}\left(\frac{b}{a}\right)& \frac{b}{a}\ll \frac{x_f}{6}, 
    \\ 
    \left[\frac{6}{x_f}\left(\frac{b}{a}\right)\right]^{-1}& \frac{x_f}{6}\ll
    \frac{b}{a}\ll[\lambda\Delta_I(E_\gamma)]^{-1},\\ 
    \frac{x_f}{6}\lambda\Delta_I(E_\gamma)&
   \frac{b}{a}\gg[\lambda\Delta_I(E_\gamma)]^{-1}.
  \end{cases}
\end{equation}

If at freezeout there are co-annihilations that increase the effective dark matter annihilation cross section, they will contribute further suppression factors to the cross section today, and thus to the mean intensity spectrum.

We conclude that, although the p-wave component does not contribute strongly to the mean intensity of the extragalactic annihilation gamma rays unless $b/a\agt10^6$, even mild values of $b$ affect the relic density constraint, resulting in significant reductions to the intensity today.  

Of course, the relic density constraint does not apply to non-thermal
relics, such as the post-freezeout reheating scenario \cite{Dutta:2009uf}.

\section{Example p-wave effects in supersymmetric extensions of the standard model}

\subsection{P-wave annihilation in the MSSM}
\label{sec:MSSM}

In Section~\ref{sec:inten2ndterm}, we justified the intuitive notion that, in order for p-wave annihilation effects to be important when calculating the mean intensity spectrum of extragalactic dark matter annihilation photons, one needs $b/a\agt10^6$.  In Section~\ref{sec:intenrd}, we argued that this is true, so long as the correct associated s-wave annihilation component needed to satisfy the relic density constraint is used.

In the case of the MSSM, we find model parameters that satisfy the relic density constraint and determine the s-wave and p-wave components of the dark matter relative-velocity-weighted annihilation cross section for that theory \cite{Drees:1992am}.  It is interesting to consider which regions of the MSSM have the largest ratio of p-wave to s-wave components.

The first step is recognizing that neutralino dark matter has certain annihilation channels where the s-wave components are helicity-suppressed.  For annihilation into a fermion anti-fermion pair, the s-wave annihilation component is suppressed by $(m_f/m_{\tilde{\chi}_1^0})^2$, the squared ratio of the fermion mass to the neutralino mass.  This includes t and u-channel sfermion exchange, and s-channel mediation by the $Z$ boson or neutral Higgs bosons.  Each contribution to the total cross-section amplitude has an approximate factor of $(m_M/m_{\tilde{\chi}_1^0})^{-2}$, where $m_M$ is the mediator mass.  This factor enhances the channels mediated by the $Z$ and lightest Higgs, but suppresses the sfermion and heavy Higgs channels.

In the parameter space where the neutralino is very nearly pure bino and the magnitude of the Higgs superpotential coupling $\mu$ is much larger than the first soft gaugino mass $M_1$, the annihilation is dominated by the sfermion exchange.  This allows us to greatly reduce the contributions of $Z$ or Higgs mediation, as well as all other annihilation channels that are not helicity-suppressed.

In the case that the sfermion masses are unified at the GUT scale, the heavy third generation fermion channels dominate the s-wave annihilations, due to their larger masses compared with the other fermions.  In parameter space where the neutralino is nearly pure bino, squark masses are much larger than sfermion masses.  However, the $b$ quark is sufficiently massive when compared to the $\tau$ lepton that it still has a significant branching ratio.  Also, remaining modest amounts of $Z$ mediation will add to the $b\overline{b}$ production.  The production of $t$ quarks becomes important if the neutralino is massive enough to kinematically allow it.

We can increase $b/a$ further by taking advantage of the sfermion mass suppression of the cross section and considering large non-universal 3rd generation sfermion masses at the GUT scale.  In this parameter space of the MSSM, where the neutralino is nearly pure bino and annihilation into 3rd generation fermions is suppressed, we would expect the dominant contribution to the s-wave component of the cross section to be proportional to $(m_\mu/m_{\tilde{\chi}_1^0})^2\alt10^{-6}$, relative to the p-wave component.

However, there are loop processes that become dominant at this point, generated when the fermion anti-fermion pair close the loop and two gauge bosons come off the internal lines \cite{Bergstrom:1997fh}, such as two photons, photon and $Z$, or two gluons.  The amplitude due to these loops provides a hard lower bound on the s-wave component of the cross section on the order of $\unit{10^{-29}}{\centi\meter\cubed\per\second}$, keeping $b/a\alt10^4$ in the MSSM parameter space that satisfies the relic density constraint.  Correspondingly, the p-wave intensity term will have a magnitude of less than a percent of the s-wave approximation.  But as we explained in Section~\ref{sec:intenrd}, such large values of $b/a$ require a significant reduction in the s-wave component of the cross section in order to satisfy the relic density constraint.

Therefore we conclude that, in the MSSM, the s-wave approximation of the intensity calculation in Eq.~(\ref{eq:S-Intensity}) gives very accurate results in the MSSM, as long as the correct annihilation cross section is used.  We find that p-wave suppression of the s-wave component is very significant in parts of the MSSM parameter space.  The intensity is further suppressed in parameter space with significant co-annihilations at dark matter freezeout.

As an example, we showed in Figure~\ref{fig:tanb10example} a comparison of the predicted extragalactic annihilation spectrum for three different MSSM models in the mSUGRA parameter space.  We expect conclusions about all other regions to follow similarly from the results of this section, understood from Figure~\ref{subfig:pwavesupplog}.  Although we see slight variations in the models' spectral shape due to differences in the photon spectrum per annihilation in the different regions, the dominant effect is the normalization of extragalactic intensity.  The co-annihilation region spectrum is so strongly p-wave suppressed (and also mildly suppressed further because of the co-annihilation factors) that it is a factor of $\mathcal{O}(10^{-2})$ less intense than the corresponding focus point/hyperbolic region spectrum that has nearly no p-wave or co-annihilation suppression.

The large p-wave strength of $b/a=378.8$ of the co-annihilation region can be accounted for by the fact that the lightest neutralino is about 99\% bino and 1\% higgsino.  Therefore, its annihilation is dominated by helicity-suppressed processes: 66\% into $b\overline{b}$ and 24\% into $\tau^+\tau^-$.  In contrast, the focus point/hyperbolic region's lightest neutralino is 72\% bino, 26\% higgsino, and 2\% wino.  It annihilates predominantly into $W^+W^-$ (54\%), $b\overline{b}$ (20\%), and $ZZ$ (18\%).

Note that not all co-annihilation regions of mSUGRA are strongly p-wave suppressed.  At larger $\tan\beta$, the s-wave annihilation into fermions mediated by an s-channel pseudoscalar Higgs is enhanced, producing mostly $b \overline{b}$ pairs, and the p-wave strength decreases back down to $\mathcal{O}(1)$ in the co-annihilation region.  For example, the upper curve in Figure~\ref{fig:bmlfig} shows the intensity spectrum for a co-annihilation region mSUGRA model as before, except $\tan\beta=50$ and $m_{\tilde{\chi}_1^0}=\unit{550}{\giga\electronvolt}$.  In this model, the p-wave strength of $b/a=4.8$ is lower because of the contribution of the pseudoscalar-mediated annihilation channel, even though the lightest neutralino is 99.8\% bino.  Near the edge of the current $b\rightarrow s\gamma$ bounds on the co-annihilation region parameter space, where $m_{\tilde{\chi}_1^0}=\unit{183}{\giga\electronvolt}$, the p-wave strength is even lower at $b/a=2.7$.  The p-wave strength increases for higher neutralino mass in this co-annihilation region because: $\mu$ increases rapidly with neutralino mass, and the pseudoscalar mass increases with $\mu$; therefore, the s-wave annihilation with pseudoscalar mediation becomes important.

\subsection{A p-wave dominated scenario}
\label{sec:pdominated}

\begin{figure}[t]
  \subfloat[]{\label{subfig:bmlinten}\includegraphics[width=0.47\textwidth]{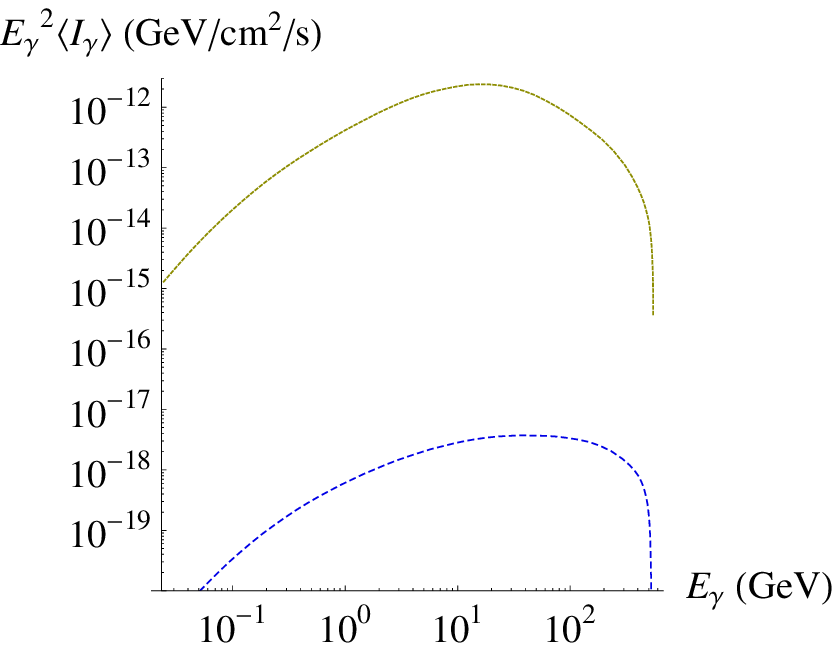}}\\
  \subfloat[]{\label{subfig:bmlratio}\includegraphics[width=0.47\textwidth]{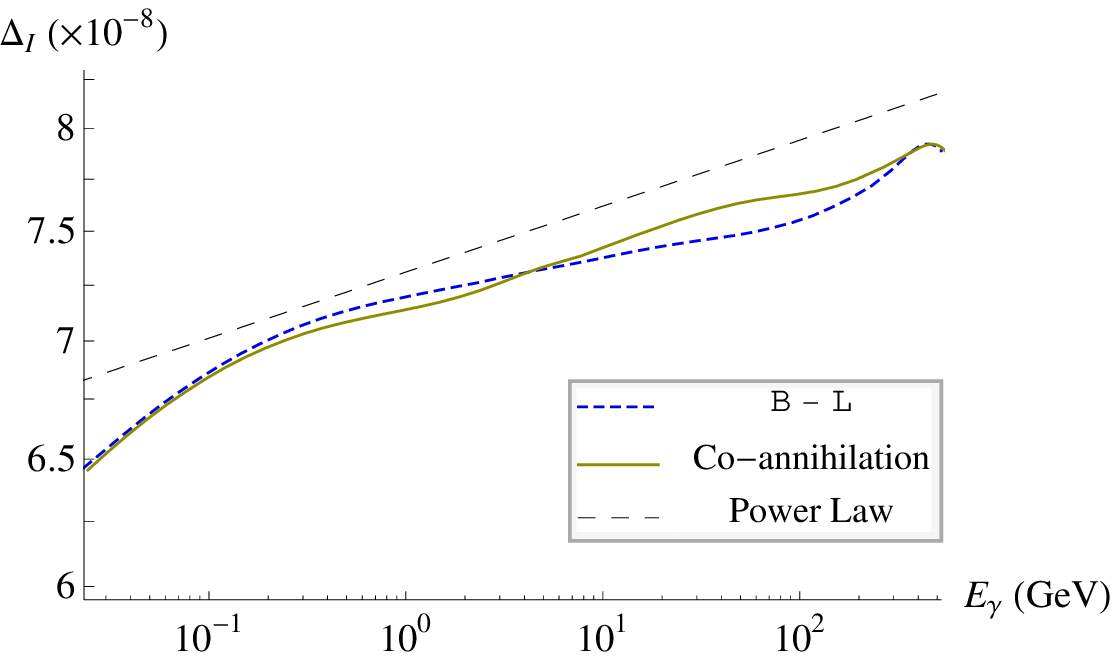}}
  \caption{\label{fig:bmlfig}\fulljust Top: The lower curve is the intensity spectrum for a pure p-wave process.  This scenario is realized in the MSSM$\otimes U(1)_{B-L}$ with right handed sneutrino dark matter.  The sneutrino, here of mass $\unit{550}{\giga\electronvolt}$, annihilates through a $Z'$ resonance into fermion anti-fermion pairs.  The $Z'$ for this plot had mass $\unit{1300}{\giga\electronvolt}$ and width $\unit{17}{\giga\electronvolt}$.  For contrast, the upper curve is due to annihilations of a $\unit{550}{\giga\electronvolt}$ neutralino in the co-annihilation region of mSUGRA with $\tan\beta=50$, $A_0=0$, and $\mu>0$.  This model has a relatively mild p-wave strength of $b/a=4.8$.  Bottom: The associated $\Delta_I$ for the two models, given by Eq.~(\ref{eq:DeltaI}).  For reference, the same power law as in Figure~\ref{subfig:DeltaI} is also shown. \hfill\mbox{}}
\end{figure}

Can we find an extension of the MSSM which would yield a larger value of $b/a$?

One interesting extension of the MSSM is that of adding a $U(1)_{B-L}$ gauge symmetry, based on a charge of baryon number minus lepton number \cite{Mohapatra:1980qe,*[{[Erratum-ibid. }][{]}]Mohapatra:1980:qf}.  This model is interesting because it provides a gauge structure that accommodates the right-handed neutrino.

In this model, the lightest supersymmetric particle (LSP) can be the lightest right sneutrino or the lightest neutralino.  Here, we consider a parameter space where the right sneutrino is the LSP.  Thus it is neutral to standard model charges and the only gauge interaction is with the $Z'$ (and $B-L$ neutralinos) via its lepton charge.  It also interacts with the $B-L$ Higgs fields via the $D$-term.  Possible s-wave annihilation processes are annihilation into neutrinos mediated by $B-L$ neutralinos, and into $B-L$ Higgs via s-channel $Z'$ or Higgs fields.  However, there is parameter space where the $B-L$ Higgs' and neutralinos are massive compared to twice the sneutrino mass, and can be neglected in this discussion.  In this region, sneutrino annihilation is (at tree level) exclusively s-channel via the $Z'$ into fermion anti-fermion pairs.  In this process, s-wave annihilation is completely forbidden.  So, this is an example of a pure p-wave annihilation process.

At one loop, an s-wave component is generated, but is strongly coupling-suppressed when compared to the tree level p-wave cross section, with additional factors of $\alpha^4$ or $g'^4\alpha^2$, where $\alpha$ is the fine structure constant (or the strong force constant in the case where gluons are emitted from quarks, instead of photons) and $g'$ is the $U(1)_{B-L}$ gauge coupling, which we take to be 0.4.  Thus, we would estimate $b/a$ in this scenario to be $\gtrsim10^8$, completely p-wave dominated.  We carry out a calculation of the intensity spectrum due to annihilations at the tree level, neglecting the small s-wave component altogether.

For our example, we consider a model with spectrum $m_{\tilde{\nu}}=\unit{550}{\giga\electronvolt}$, $M_{Z'}=\unit{1300}{\giga\electronvolt}$, and $Z'$ width $\Gamma_{Z'}=\unit{17}{\giga\electronvolt}$.  Here, the sneutrino is at a near resonance with the $Z'$, which allows the relic density constraint to be satisfied.  The photon spectrum per annihilation was simulated with Pythia 8.135 \cite{Sjostrand:2007gs}.  The resulting intensity spectrum is shown in the bottom curve of Figure~\ref{fig:bmlfig}.  The low intensity in this model shows the expected full p-wave suppression of $10^6$, as described in Section~\ref{sec:intenrd}.  Here, the differences in spectrum shape between the two models will be due to both differences in photon spectrum per annihilation, and the fact that the lower intensity curve is directly coupled to the current dark matter velocity distribution, whereas the upper curve is not.

We note that in the context of the $B-L$ extension of the MSSM, there are also parameter regions with significant s-wave annihilation components.  For example, regions with light $B-L$ Higgs fields open up sneutrino annihilation channels involving only scalar fields whose contribution to the cross section is momentum-independent \cite{Allahverdi:2009ae}.

\section{Intensity spectra of extragalactic annihilation gamma rays for general velocity-dependent particle models}
\label{sec:generalv}

Adding a p-wave term to the annihilation cross section Eq.~(\ref{eq:cross section}) is sufficient to describe the physics at energies of the current dark matter distribution for many particle physics models.  However, there is a variety of additional momentum-dependencies that can arise.

It is therefore useful to develop a formalism for calculating mean extragalactic annihilation photon intensity spectra for general velocity-dependent particle physics models.  Given a relative-velocity-weighted annihilation cross section as a function of squared relative velocity $[\sigma v](v^2)$, the mean photon intensity due to extragalactic annihilations is simply
\begin{equation}
  \label{eq:genI}
  \mean{I_\gamma}(E_\gamma)=\int\frac{\der z}{H(z)}\mean{\delr^2\frac{[\sigma v](v^2)}{[\sigma v]_0}}\!(z)\,W((1+z)E_\gamma,z),
\end{equation}
where, again, $[\sigma v]_0$ is any convenient reference cross section used in the intensity window function $W$, in Eq.~(\ref{eq:window}).

To calculate the ensemble average, we use the following result from the spherical halo model.  Let $X_1(\mathbf{r},z),\dots,X_N(\mathbf{r},z)$ be any fields associated with universal halo functions $X_{1h}(r|M,z),\dots,X_{Nh}(r|M,z)$.  The overdensity of $X_i$ is
\begin{equation}
  \label{eq:overdensityX}
  \delta_{X_i}(\mathbf{r},z)=\frac{X_i(\mathbf{r},z)}{\overline{X_i}(z)}-1
\end{equation}
where
\begin{equation}
  \label{eq:meanX}
  \overline{X_i}(z)=\int\der M\frac{\der n}{\der M}(M,z)\int\der^3\mathbf{r}X_{ih}(r|M,z).
\end{equation}
The ensemble average of any product of these overdensities, all evaluated at one position, at redshift $z$ is
\begin{equation}
  \mean{\prod_{i=1}^N\delta_{X_i}}\!(z)=\int\der M\frac{\der n}{\der M}(M,z)\int\der^3\mathbf{r}\prod_{i=1}^N\frac{X_{ih}(r|M,z)}{\overline{X_i}(z)},
\end{equation}
where we have used the disjoint, rigid halo approximation \footnote{This expression is planned to be justified in an upcoming paper}.  Using Taylor's theorem, we find a formula for the ensemble average for any piecewise analytic function $F$ of the overdensities of quantities associated with universal halo functions.
\begin{eqnarray}
\nonumber
& &  \Bigl\langle F(\delta_{X_1},\dots,\delta_{X_N})\Bigr\rangle(z)\\
\nonumber
&=&\int\der M\frac{\der n}{\der M}(M,z)\\
& &\times \int\der^3\mathbf{r}\;F\!\left(\frac{X_{1h}(r|M,z)}{\overline{X_1}(z)},\dots,\frac{X_{Nh}(r|M,z)}{\overline{X_N}(z)}\right).
\end{eqnarray}

In the case of Eq.~(\ref{eq:genI}), we re-express $\sigma v$ as a function of $\delvs$ at redshift $z$ by substituting $v^2=\lambda\overline{\sigma_v^2}(z)\bigl(1+\delvs\bigr)$.  We then obtain
\begin{eqnarray}
\nonumber
  \mean{\delr^2\frac{[\sigma v](\delvs)}{[\sigma v]_0}}&&(z)=\int\der M\frac{\der n}{\der M}(M,z)\\
& &\hspace{-20pt}\times\int\der^3\mathbf{r}\frac{\rho_h^2(r|M,z)}{\overline{\rho}^2(z)}\frac{[\sigma v]\!\!\left(\frac{\sigma_{vh}^2(r|M,z)}{\overline{\sigma_v^2}(z)}\right)}{[\sigma v]_0}.
  \label{eq:genvmoment}
\end{eqnarray}
There are calculations where it may be more convenient just to, when possible, treat the cross section itself as a universal halo function, given by Eq.~(\ref{eq:halo cs}).  In those situations, one can define the mean cross section $\overline{\sigma v}$ at redshift $z$ and associated overdensity $\delta_{\sigma v}$ field in the usual way, with Eq.~(\ref{eq:overdensityX}) and Eq.~(\ref{eq:meanX}).  Then we have the alternative expression
\begin{eqnarray}
\nonumber
 \mean{\delr^2\frac{[\sigma v]}{[\sigma v]_0}}(z)
&=&\frac{\overline{\sigma v}(z)}{[\sigma
v]_0}\Bigl\langle\delr^2(1+\delta_{\sigma v})\Bigr\rangle(z)\\
\nonumber
  &=&\frac{\overline{\sigma v}(z)}{[\sigma v]_0}\int\der M\frac{\der
   n}{\der M}(M,z)\\
\nonumber
& &\times\int\der^3\mathbf{r}\frac{\rho_h^2(r|M,z)}{\overline{\rho}^2(z)}\left(1+\frac{[\sigma v]_h(r|M,z)}{\overline{\sigma v}(z)}\right).\\
\end{eqnarray}

Similarly, if the annihilation product spectrum per annihilation $\frac{\der N}{\der E}$ should vary significantly over the range of annihilation center-of-momentum energies that occur, then it may also be expressed as a function of relative velocity and included in the ensemble average.

\begin{figure*}[t]
  \subfloat[]{\label{subfig:sfeldephi}\includegraphics[width=0.47\textwidth]{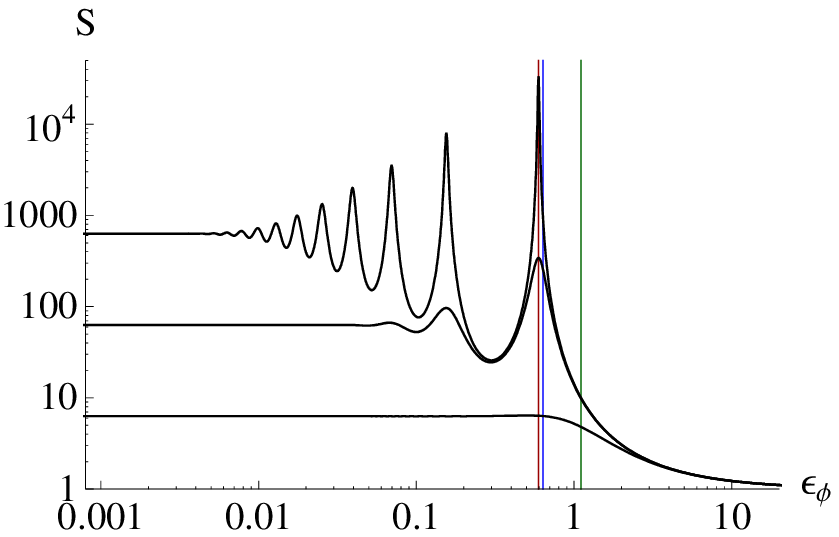}}\quad\quad
  \subfloat[]{\label{subfig:sfeldvsq}\includegraphics[width=0.47\textwidth]{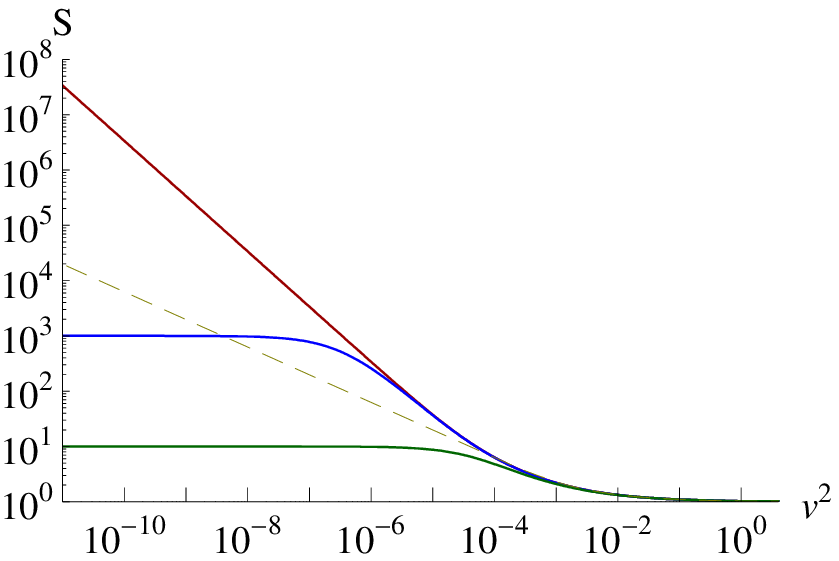}}
  \caption{\label{fig:sfeld}\fulljust Left: The s-wave Sommerfeld enhancement $S(\epsilon_v|\epsilon_\phi)$ vs. $\epsilon_\phi$ for $\epsilon_v=1$, $0.1$, and $0.01$ from bottom to top, respectively.  The vertical lines correspond, from right to left, to $\epsilon_\phi=1.107$, $0.635$, and the first Sommerfeld resonance $\epsilon_\phi^{^{(1)}}$ described in the text.  Right: $S(\frac{v}{\alpha}|\epsilon_\phi)$ vs. $v^2$ for $\alpha=0.01$. The solid curves show the enhancement for the same three values of $\epsilon_\phi$ specified earlier from bottom to top, and the dashed curve shows the Coulomb case where $\epsilon_\phi=0$. \hfill\mbox{}}
\end{figure*}

\section{Sommerfeld enhancement and resonance effects in the mean intensity spectrum}
\label{sec:sommerfeld}

The Sommerfeld enhancement \cite{Hisano:2004ds,*ArkaniHamed:2008qn, *Lattanzi:2008qa, *MarchRussell:2008tu, Iengo:2009ni, Iengo:2009xf, *Cassel:2009wt} of dark matter annihilation occurs in cases where the annihilation is mediated by an attractive Yukawa force through a scalar or vector boson.  Non-perturbative resonant boson exchange between annihilating particles is demonstrated to result in a significant enhancement of the cross-section that grows as relative particle motion decreases.  

Although this enhancement occurs for any partial wave \cite{Iengo:2009ni, Iengo:2009xf, *Cassel:2009wt}, we will (for simplicity) present intensity spectra for s-wave Sommerfeld enhancement.  It is expressed in the form
\begin{equation}
  [\sigma v](v) = S\left(\frac{v}{\alpha}\Big|\epsilon_\phi\right)[\sigma v]_0
\end{equation}
where $[\sigma v]_0$ is the relative-velocity-weighted annihilation cross section at tree level, assumed constant at all relevant energies for this model, and set for our calculations to $[\sigma v]_f=\unit{3\times10^{26}}{\centi\meter\cubed\per\second}$ to satisfy the relic density (although, in careful calculations, the enhancement effect can have some affect on the relic density calculation \cite{Feng:2010zp}). Here, $\alpha$ is the Yukawa coupling between the dark matter and mediator, and 
\begin{equation*}
  \epsilon_\phi\equiv\frac{m_\phi}{m_X\alpha}
\end{equation*}
with $m_\phi$ the mediator mass and $m_X$ the dark matter mass.  For convenience, we also define $\epsilon_v\equiv\frac{v}{\alpha}$.  

Following the derivation by \cite{Iengo:2009ni}, $S$ is extracted from the solution $\Phi(x)$ of the Schr\"{o}dinger equation
\begin{equation*}
  \Phi''+\frac{2}{x}\Phi'+(1-\tilde{U})\Phi=0
\end{equation*}
with boundary conditions $\Phi(0)=1$ and $\Phi'(0)=-1/\epsilon_v$, and where
\begin{equation*}
  \tilde{U}(x)\equiv-\frac{2}{\epsilon_vx}e^{-2\frac{\epsilon_\phi}{\epsilon_v}x}
\end{equation*}
is the normalized potential.  This is more illuminating with $\psi(x)\equiv x\Phi(x)$, in which case the Schr\"{o}dinger equation becomes
\begin{equation}
  \label{eq:someq}
  \psi''+(1-\tilde{U})\psi=0
\end{equation}
with the necessary solution near the boundary of $\lim_{x\rightarrow0}\psi=x-x^2/\epsilon_v$ and $\lim_{x\rightarrow0}\psi'=1-2x/\epsilon_v$.  

It is now easy to see that $\psi$ converges very quickly to a sinusoid as $x$ increases.  $S$ is simply the inverse square of the amplitude of $\psi$ far from the origin.  One can integrate Eq.~(\ref{eq:someq}) to $x=x_M$ large enough that $\tilde{U}(x_M)$ is sufficiently negligible and simply evaluate
\begin{equation}
  S=\frac{1}{\psi^2(x_M)+\psi'^2(x_M)}.
\end{equation}

Figure~\ref{subfig:sfeldephi} shows the enhancement for $\epsilon_v=1$,
$0.1$, and $0.01$.  As relative velocity decreases, a series of
Sommerfeld resonances reveals itself.  
Let us define the locations of the resonances to be at $\epsilon_\phi=\epsilon_\phi^{^{(n)}}$ for $n=1,2,\dots.$  Analytic approximations show the first few s-wave resonances to be near $\epsilon_\phi^{^{(n)}}\approx6/(n\pi)^2$.  Also, 
\begin{equation*}
  \lim_{\epsilon_v\rightarrow0}S(\epsilon_v|\epsilon_\phi^{^{(n)}})=\frac{A^{^{(n)}}}{\epsilon_v^2}
\end{equation*}
where $A^{^{(n)}}$ are constants.  We find the first resonance at $\epsilon_\phi^{^{(1)}}=0.595\,306\,210\,530\,309$ to have $A^{^{(1)}}=3.37286$ up to the given precisions.  Away from the resonances, the enhancement saturates to a constant value as $v$ diminishes.  Until the resonances arise at low $\epsilon_\phi$, $S$ follows the Coulomb-Sommerfeld enhancement in the center-of-momentum frame
\begin{equation*}
  S(\epsilon_v|0)=\frac{2\pi/\epsilon_v}{1-e^{-2\pi/\epsilon_v}}
\end{equation*}
which goes like $\epsilon_v^{-1}$ at low $v$.  Examples of the relative velocity dependence of each of these cases are shown in Figure~\ref{subfig:sfeldvsq} for $\alpha=0.01$.

\begin{figure*}
  \subfloat{\label{subfig:sfeldintenee}\includegraphics[width=0.47\textwidth]{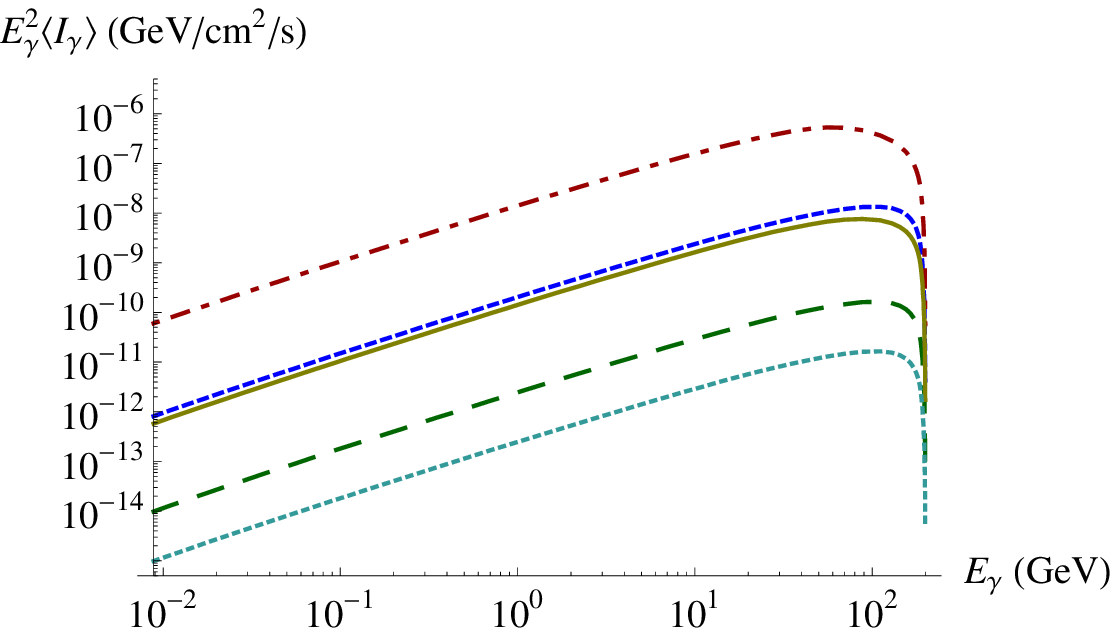}}
  \subfloat{\label{subfig:sfeldintentautau}\includegraphics[width=0.47\textwidth]{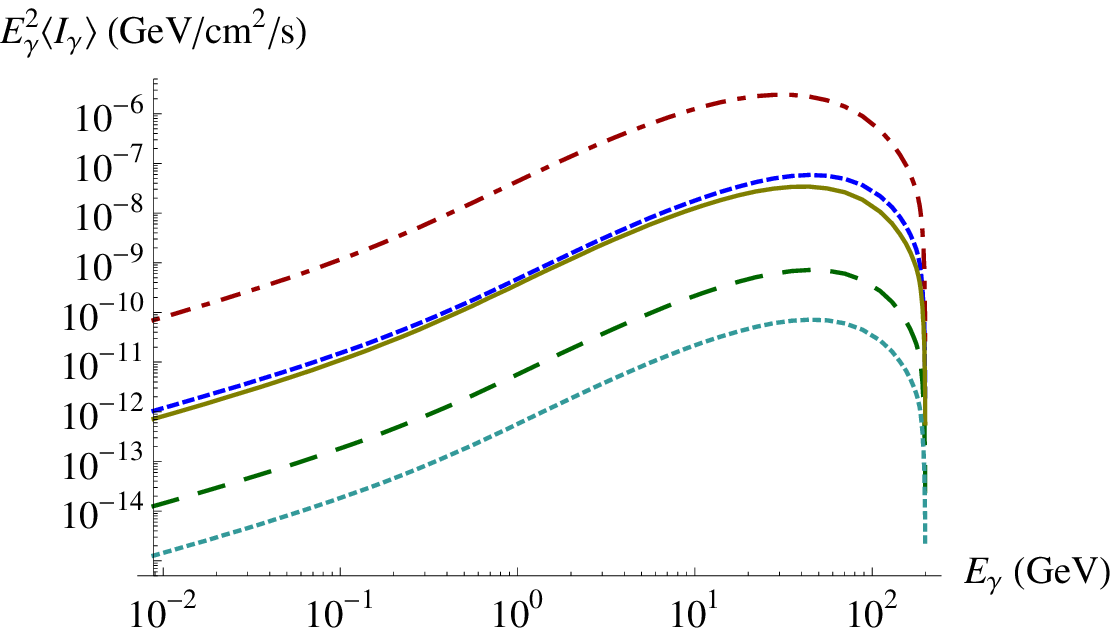}}
  \caption{\label{fig:sfeldinten}\fulljust Extragalactic annihilation gamma-ray intensities for s-wave theories, sample Sommerfeld-enhanced s-wave theories, and Sommerfeld s-wave resonances.  For these models, we set $\alpha=0.01$ and $m_X=\unit{200}{\giga\electronvolt}$, and used $\lambda=6$ when generating $[\sigma v](\delvs)$ in Eq.~(\ref{eq:genvmoment}).  The bottom dotted curves show the intensity for no Sommerfeld enhancement, the solid lines show the Coulomb-Sommerfeld resonance, the top dot-dashed curves show the first Sommerfeld resonance, and the long (short) dashed curves show the Sommerfeld enhanced intensity where the enhancement saturates at 10 (1000).  Left: primary photon radiation from annihilation into electron-positron pairs.  Right: annihilation into $\tau^+ \tau^-$. \hfill\mbox{}}
\end{figure*}

\begin{figure*}
  \subfloat{\label{subfig:sfeldratioee}\includegraphics[width=0.37\textwidth]{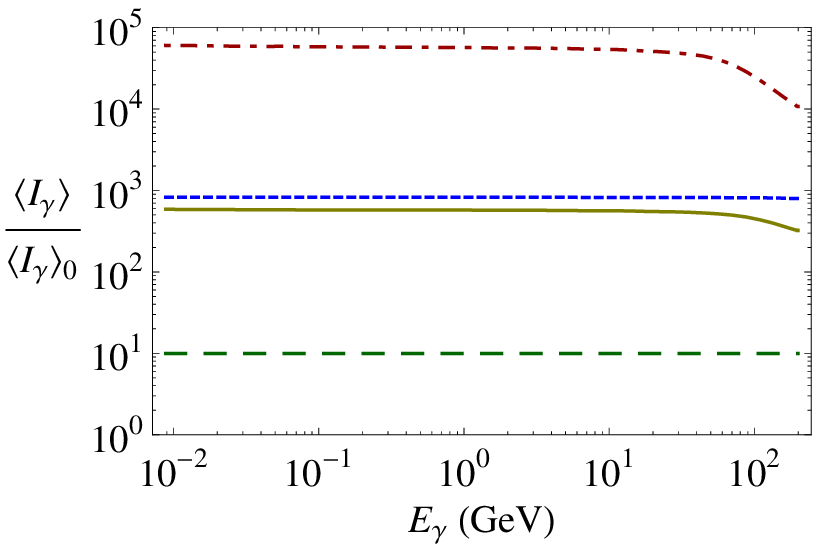}}
  \quad
  \subfloat{\includegraphics[width=0.13\textwidth]{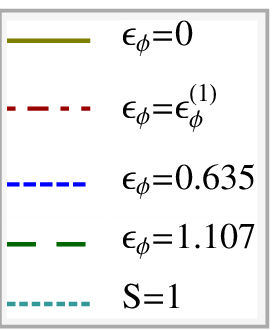}}
  \quad
  \subfloat{\label{subfig:sfeldratiotautau}\includegraphics[width=0.37\textwidth]{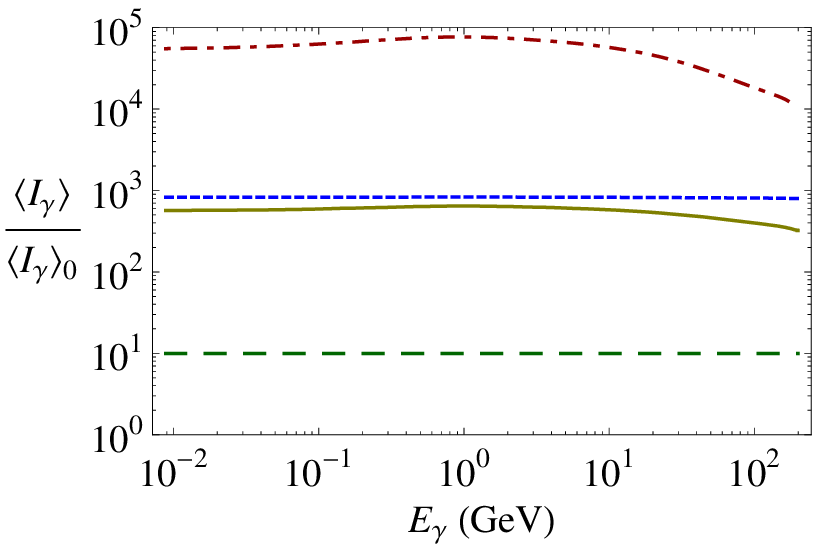}}
  \quad\quad\mbox{}
  \caption{\label{fig:sfeldinten2}\fulljust Ratios of the predicted Sommerfeld enhanced intensities to the unenhanced intensity. Left: primary photon production from annihilation exclusively into electron-positron pairs. Right: annihilation into $\tau^+ \tau^-$.\hfill\mbox{}
}
\end{figure*}

\begin{figure*}
  \subfloat[$\epsilon_\phi=1.107$]{\label{subfig:somrat1}\includegraphics[width=0.3\textwidth]{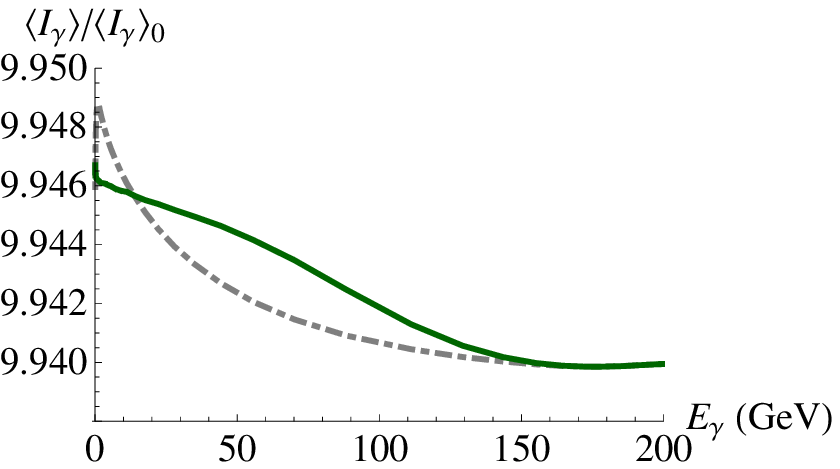}}
  \subfloat[$\epsilon_\phi=0.635$]{\label{subfig:somrat2}\includegraphics[width=0.3\textwidth]{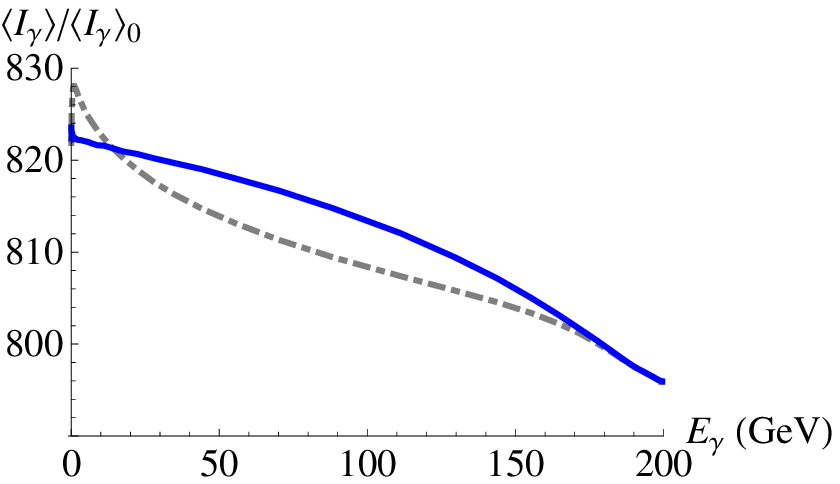}}\\
  \subfloat[$\epsilon_\phi=0$]{\label{subfig:somrat3}\includegraphics[width=0.3\textwidth]{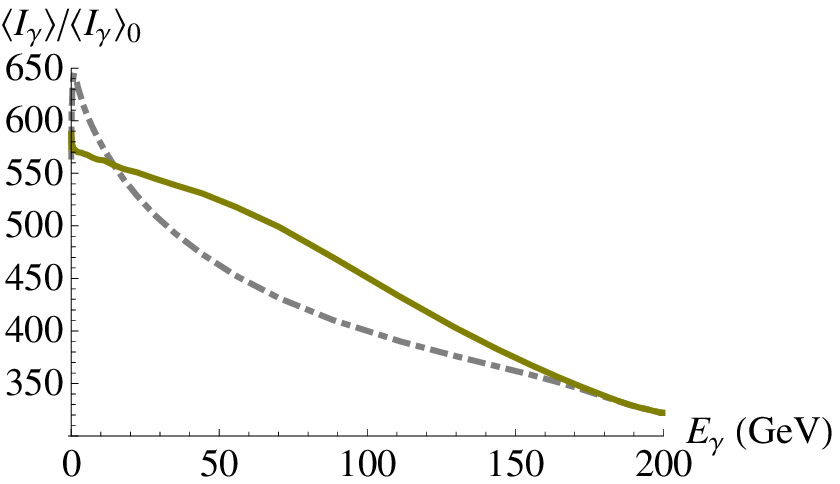}}
  \subfloat[$\epsilon_\phi=\epsilon_\phi^{^{(1)}}$] {\label{subfig:somrat4}\includegraphics[width=0.3\textwidth]{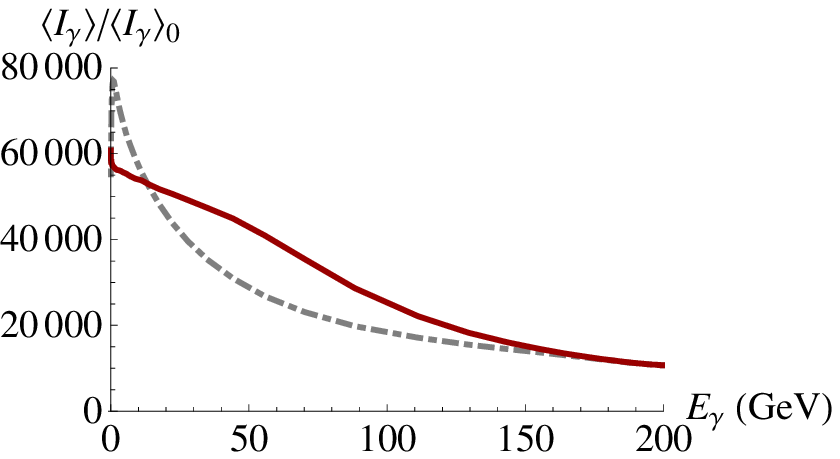}}  
  \caption{\label{fig:sfeldratiolin}\fulljust The intensity ratios in Figure~\ref{fig:sfeldinten2}.  The solid curves are for annihilation into $e^+ e^-$, and the dot-dashed curves show the results for $\tau$ pair production.  \hfill\mbox{}}
\end{figure*}

Figure~\ref{fig:sfeldinten} shows the results for the extragalactic gamma-ray intensity due to annihilations for theories with these cross sections.  Again for simplicity, we consider theories that annihilate exclusively into lepton anti-lepton pairs.  The photon radiation spectra per annihilation for these processes were also simulated with Pythia. In these examples, the dark matter mass is taken to be $m_X=\unit{200}{\giga\electronvolt}$.  Higher masses were explored; they simply gave the same results, scaled up in energy.  Annihilation into $\mu^+ \mu^-$ is visually indistinguishable from the electron case, with a slightly higher intensity.  The ratio of each enhanced case to its respective s-wave approximation (where $S=1$) is shown in Figure~\ref{fig:sfeldinten2}.  Here we find that, for $\alpha=0.01$, the intensity enhancement is nearly uniform over most of the Sommerfeld parameter space.  This follows from the fact that, as we see in Figure~\ref{subfig:sfeldvsq}, the cross section has already saturated at the relative velocities important today, around $10^{-4}$, unless you are very, very close to a resonance value of $\epsilon_\phi$.  However, for smaller values of $\alpha$, the graph in Figure~\ref{subfig:sfeldvsq} shifts to the left and the cross section may not necessarily be completely saturated today for larger enhancements.  This results in a smaller cross section at high energies and has the effect of widening the intensity peak and shifting the maximum to smaller energy.  This is what is observed with the resonance cases: $\epsilon_\phi=0$ (an example of a $v^{-1}$ resonance), and $\epsilon_\phi=\epsilon_\phi^{^{(n)}}$ ($v^{-2}$ resonances).  

To see the detail of the variation of $\mean{I_\gamma}/\mean{I_\gamma}_0$ at peak intensities, plots of each ratio on a blown up linear scale are shown in Figure~\ref{fig:sfeldratiolin}.  For comparison, the enhancement at saturation for $\epsilon_\phi=1.107$ was 10.00, and for $\epsilon_\phi=0.635$ it was 1004.  The intensity ratio for annihilation into taus in the saturated examples was very similar to the electron-production results.  However, we can see that differences in the spectrum per annihilation become important at the resonances.  

In principle, the resonance cross-sections can break unitarity bounds for s-wave annihilation \cite{Backovic:2009rw}.  If the scattering operator conserves angular momentum and is unitary, then the weighted s-wave annihilation cross section must satisfy \cite{Hui:2001wy}
\begin{equation}
  \sigma v\leq\frac{4\pi}{m_X^2v}.
\end{equation}
In the low-$v$ limit $v\ll\alpha$ for the Coulomb case, this provides an upper bound on the mass for a given coupling:
\begin{equation}
  \alpha m_X^2\lesssim\frac{2}{[\sigma v]_f}\sim(\unit{30}{TeV})^2,\ \ \text{for }\epsilon_\phi=0
\end{equation}
for the value of $[\sigma v]_f$ in our model.  For the first Sommerfeld resonance, we require
\begin{equation}
  v\gtrsim v_{\text{min}}\equiv\frac{A^{^{(1)}}m_X^2\alpha^2}{4\pi}[\sigma v]_f\sim\left(\frac{m_X\alpha}{\unit{40}{TeV}}\right)^2,\ \ \text{for }\epsilon_\phi=\epsilon_\phi^{^{(1)}}.
\end{equation}
As long as the enhancement saturates before getting to relative velocities below $v_{\text{min}}$, then the theory is consistent.  Otherwise, some neglected model-dependent effects (such as finite widths or non-perturbative dynamics) become important and must be taken into account.  A model near resonance that saturates below scales that contribute to the intensity would be indistinguishable from the resonance intensity, but would still satisfy unitarity constraints.

We finish this section by mentioning another common example of resonance effect: the Breit-Wigner resonance due, for example, to the mass of the mediator being at the energy of the annihilating particles.  At non-relativistic speeds, this phenomenon also results in saturated cross sections, but in certain situations such as when the dark matter mass is very near half the resonance mass and much larger than the resonance width, we have $\sigma v\sim v^{-4}$ for non-relativistic relative velocities above the saturation scale (see Appendix~\ref{ap:bwres}).

\section{Discussion}
\label{sec:discussion}

In this paper, we have studied how velocity-dependent annihilation of dark matter particles can affect the mean intensity spectrum of primarily produced,
extragalactic, annihilation gamma-ray background.  Velocity-dependence of dark matter annihilation contains information about the spin of the dark matter and its annihilation mediators, and about various possible resonance effects.

We first considered the effects of p-wave annihilation, where the annihilation cross section is well approximated by $\sigma v=a+bv^2$. This is an important feature in minimal supersymmetric extensions of the standard model where the s-wave component can be strongly helicity-suppressed.  If the relative-velocity-weighted cross section is quadratic in relative velocity at all energy scales up to freezeout, then the relic density constraint requires the s-wave component of the cross section to be reduced for larger p-wave components, resulting in a suppression of the extragalactic gamma-ray intensity today.

Other than setting the scale of the s-wave annihilation component, we find that the p-wave does not contribute to the intensity unless the coefficient $b$ is at least 6 orders of magnitude greater than the s-wave component.  We do not find such a scenario within the MSSM (we found MSSM p-waves only as high as 4 orders of magnitude greater than the s-wave component); therefore, differences in mean intensities in the MSSM are due to factors from different s-wave components (from the presence of p-wave annihilation or co-annihilations), or major differences in the photon spectrum per annihilation.

Although there was no model found in the MSSM where p-wave annihilation dominates today (with $b/a\agt10^6$), we presented such a scenario within the context of MSSM$\otimes U(1)_{B-L}$, where baryon number minus lepton number is an additional gauge charge with new $Z'$ vector boson.  This model contains sneutrino dark matter at a near resonance with the $Z'$ in order to satisfy the relic density constraint, but the cross section remains proportional to the squared relative velocity at tree level.  In this case, the intensity spectrum is maximally coupled to the velocity distribution, but the dominant feature is again the strong $10^{-6}$ suppression in the normalization of the intensity curve.

However, one cannot conclude that theories with strong p-wave annihilation will always result in an unobservably small extragalactic gamma-ray signal, since non-thermal freezeout of the dark matter can allow for larger intensities.

A possible cross section feature that we found to enhance the mean intensity of extragalactic annihilation gamma rays was the Sommerfeld-enhanced s-wave cross section.  If the dark matter attracts itself, then s-wave annihilation becomes stronger at low energies.  Again, the dominant feature is a change in the normalization of the mean intensity $\mean{I_\gamma}(E_\gamma)$, but this time an increase by as high as $10^4$.  The Sommerfeld enhancement also has the effect of shifting the peak of $E_\gamma^2\mean{I_\gamma}$ to lower photon energy $E_\gamma$, thereby broadening the intensity peak.  The Sommerfeld resonances provide an example of $v^{-2}$ behavior for the velocity-weighted cross section, or $v^{-1}$ behavior for an extremely light or massless mediator.  An example of $v^{-4}$ behavior can be found in Breit-Wigner resonances for certain ranges of $v$.

In all of the cases for which we calculated intensities and compared them to an associated constant $\sigma v$ model, we find that the dominant effect is a change in the normalization of the mean intensity. By considering velocity effects alone, we could account for modification in intensity magnitude over 10 orders of magnitude (up to 6 orders decrease from p-wave effects, and over 4 orders increase possible from Sommerfeld enhancements).  Non-thermally produced dark matter relics complicate this even further.  

It is straightforward to extend our formalism to the calculations of the anisotropy in the gamma-ray background \cite{Ando:2005xg,*Ando:2006cr}. Different particle physics models will have unique angular power spectral features, independent of intensity normalization \footnote{Work in progress.}. For example, in a Sommerfeld-enhanced theory, small halos have smaller virial motions and will appear slightly brighter than for a theory without enhancement.  

Future work will also need to explore the robustness of these calculations to the astrophysical uncertainties, including the universal halo functions, and distribution of halos and subhalos.

If the LHC is to detect a candidate dark matter particle, the discovery will require verification of consistency with the existing dark matter population through direct and/or indirect observations.  The methods outlined here are able to improve predictions of indirect signals for certain particle models and expand the reach of interpretations of analysis of the observed signals.  In order to be able to extract new particle physics and astrophysics information from the intensity of an indirect signal, the challenge remains to identify robust features that can be disentangled from the uncertainties of the dark matter intrinsic properties and astrophysical distribution.

\begin{acknowledgments}
SC is grateful to Shin'ichiro Ando for his kind assistance with details on the numerical calculations of the matter density distribution, and to Joachim Edsj\"{o} for offering details of the annihilation spectra in DarkSUSY.  SC and BD are supported by DOE Grant DE-FG02-95ER40917.  EK is supported in part by NSF grant PHY-0758153.
\end{acknowledgments}

\appendix
\section{Effect of Low Mass Halos}
\label{ap:smallminmass}

\begin{figure}
  \includegraphics[width=0.45\textwidth]{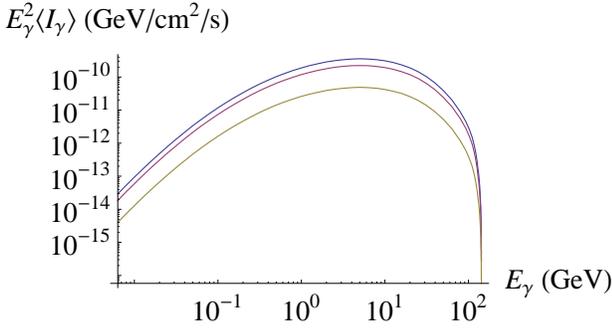}
  \caption{\label{fig:fp_minmass}\fulljust Intensity spectra of dark matter annihilation gamma-rays for the MSSM focus point model described in Section~\ref{sec:inten2ndterm}, with minimum halo mass $M_{\text{min}}=10^6 M_\odot$, $10^{-6} M_\odot$, and $10^{-12} M_\odot$ from bottom to top. \hfill\mbox{}}
\end{figure}

The intensity of photons from extragalactic annihilating dark matter is sensitive to the scale of minimal halo mass, $M_{\text{min}}$, which is dependent on details of the dark matter self-interaction.  For simplicity, the spectra calculated in this paper set $M_{\text{min}}=10^6 M_\odot$.  To gain understanding of the effect of low mass halos, we plot in Figure~\ref{fig:fp_minmass} the intensity for a model with different values of $M_{\text{min}}$.  The intensity enhancement from $M_{\text{min}}=10^6 M_\odot$, at the spectral peak, is 4.6 for $M_{\text{min}}=10^{-6} M_\odot$ and 7.3 for $M_{\text{min}}=10^{-12} M_\odot$.

\begin{figure}
  \includegraphics[width=0.45\textwidth]{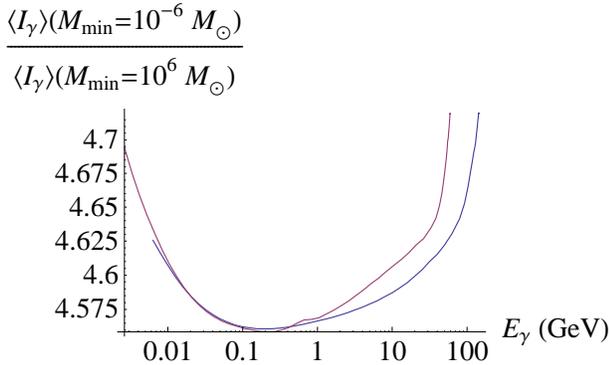}
  \caption{\label{fig:minmassrat}\fulljust Intensity enhancement factor, when decreasing the minimum halo mass from $10^6 M_\odot$ to $10^{-6} M_\odot$, for the focus point and bulk particle models described in Section~\ref{sec:inten2ndterm}.  The bulk model is the curve over slightly lower energies, since it has a less massive dark matter particle mass.}
\end{figure}

Figure~\ref{fig:minmassrat} shows the enhancement factor for $M_{\text{min}}=10^{-6} M_\odot$ (the curve offset to the right).  We see the enhancement is essentially uniform, but is slightly weaker at the spectral peak than at the tails.  For comparison, Figure~\ref{fig:minmassrat} also shows the plot for the bulk model, also described in Section~\ref{sec:inten2ndterm}.  We see that decreasing the minimum halo mass has practically the same effect for the different models.

\section{Breit-Wigner Resonances}
\label{ap:bwres}

If the dark matter annihilates into an unstable particle of mass $M$ and decay rate $\Gamma\ll M$ that then decays, and the center-of-momentum energy $E$ of the annihilation is near $M$, then the annihilation cross section is of the form \cite{Ibe:2008ye,*Peskin:1995ev,*Sakurai:1967}
\begin{equation}
  \sigma_{_{\text{BW}}}(E)\propto \frac{1}{E\sqrt{E^2-4m_X^2}}\,\frac{M^2\Gamma^2}{(E^2-M^2)^2+M^2\Gamma^2}.
\end{equation}
If we define
\begin{align*}
  \tilde{\Gamma}&\equiv \frac{\Gamma}{M},\\
  \Delta m&\equiv\frac{M}{2}-m_X,\text{ and}\\
  \overline{\Delta m}&\equiv\frac{\Delta m}{M}\left(1-\frac{\Delta m}{M}\right),
\end{align*}
then, for any relative velocity $v$ of the annihilating particles and spectral separation $\Delta m$, the velocity dependence of the cross section near the resonance is
\begin{equation}
\label{eq:BWres}
  [\sigma v]_{_{\text{BW}}}(v)\propto\frac{\sqrt{1-\left(\frac{v}{2}\right)^2}}{1+\left[\frac{\left(\frac{v}{2}\right)^2-4\overline{\Delta m}}{\tilde{\Gamma}\left[1-\left(\frac{v}{2}\right)^2\right]}\right]^2}.
\end{equation}
For a non-relativistic resonance, $v\ll1$ and $|\Delta m|\ll M$ giving
\begin{equation}
  [\sigma v]_{_{\text{BW}}}(v)=\frac{[\sigma v]_r}{1+\frac{1}{\tilde{\Gamma}^2}\left[\left(\frac{v}{2}\right)^2-4\overline{\Delta m}\right]^2}
\end{equation}
where, in this context, $\overline{\Delta m}=\Delta m/M$ and $[\sigma v]_r$ is the velocity-weighted cross section at the resonance energy where $v=v_r\equiv4\sqrt{\overline{\Delta m}}$ (which is unattainable in the case that $M<2m$).

If there are no other significant additional features in the cross section from the resonance scale $v_r$ to the freezeout scale $v_f=6T_f/m_X\sim1/2$ for freezeout temperature $T_f$, then $[\sigma v]_r$ can be correlated to the mean cross section at freezeout $[\sigma v]_f\sim\unit{3\times10^{-26}}{\centi\meter\cubed\per\second}$:
\begin{equation*}
  [\sigma v]_r\sim\frac{[\sigma v]_f}{(16\tilde{\Gamma})^2}
\end{equation*}
where we assumed $v_f\sim1/2$, $|\overline{\Delta m}|\ll1/64$, and $\tilde{\Gamma}\ll1/16$, and we neglected some $\mathcal{O}(1)$ constants due to $v_f$ being near 1.  In this kind of theory,
\begin{equation}
  [\sigma v]_{_{\text{BW}}}(v)=\frac{\eta[\sigma v]_f}{(16)^2}\,\frac{\left[1-\left(\frac{v}{2}\right)^2\right]^{5/2}}{\left[\left(\frac{v}{2}\right)^2-4\overline{\Delta m}\right]^2+\tilde{\Gamma}^2\left[1-\left(\frac{v}{2}\right)^2\right]^2}
\end{equation}
up to the freezeout scale where $\eta$ is an $\mathcal{O}(1)$ constant that takes into account our approximations of the relic density calculation and relativistic freezeout velocities.

The velocity dependence for s-wave annihilation via a non-relativistic ($v_r\ll1$ and $|\Delta m|\ll M/64$) Breit-Wigner resonance of small width ($\Gamma\ll M/16$) is therefore found to have the broad behavior of
\begin{equation}
  [\sigma v]_{_{\text{BW}}}(v)\approx
    \begin{cases}
      {\displaystyle\frac{\eta[\sigma v]_f}{(16v_s)^2}},&{\displaystyle\text{for }\frac{v}{2}\ll v_s,}\\
      \normalsize\vspace{-0.9\baselineskip}\mbox{}\\
      {\displaystyle\frac{\eta[\sigma v]_f}{16v^4}},&{\displaystyle\text{for }v_s\ll\frac{v}{2}\ll1,}
    \end{cases}
\end{equation}
where the cross section saturates at
\begin{equation}
  \frac{v}{2}\sim v_s\equiv\sqrt[4]{\tilde{\Gamma}^2+(4\overline{\Delta m})^2}.
\end{equation}
Accordingly, when the energy scale of the dark matter is above the saturation threshold, $[\sigma v]_{_\text{BW}}(v)\propto v^{-4}$, and when the cross section is saturated, it is modified by a factor of $\sim(16v_s)^{-2}$ from the freezeout cross section.  Behaviors for other cases can be similarly derived starting from Eq.~(\ref{eq:BWres}), and their features in the extragalactic gamma-ray spectrum due to dark matter annihilation can then be studied using the methods in this paper.


\begin{thebibliography}{10}%
\makeatletter
\providecommand \@ifxundefined [1]{%
 \ifx #1\undefined \expandafter \@firstoftwo
 \else \expandafter \@secondoftwo
\fi
}%
\providecommand \@ifnum [1]{%
 \ifnum #1\expandafter \@firstoftwo
 \else \expandafter \@secondoftwo
\fi
}%
\providecommand \enquote [1]{``#1''}%
\providecommand \bibnamefont  [1]{#1}%
\providecommand \bibfnamefont [1]{#1}%
\providecommand \citenamefont [1]{#1}%
\providecommand\href[0]{\@sanitize\@href}%
\providecommand\@href[1]{\endgroup\@@startlink{#1}\endgroup\@@href}%
\providecommand\@@href[1]{#1\@@endlink}%
\providecommand \@sanitize [0]{\begingroup\catcode`\&12\catcode`\#12\relax}%
\@ifxundefined \pdfoutput {\@firstoftwo}{%
 \@ifnum{\z@=\pdfoutput}{\@firstoftwo}{\@secondoftwo}%
}{%
 \providecommand\@@startlink[1]{\leavevmode}%
 \providecommand\@@endlink[0]{}%
}{%
 \providecommand\@@startlink[1]{%
  \leavevmode
  \pdfstartlink
   attr{/Border[0 0 1 ]/H/I/C[0 1 1]}%
   user{/Subtype/Link/A<</Type/Action/S/URI/URI(#1)>>}%
  \relax
 }%
 \providecommand\@@endlink[0]{\pdfendlink}%
}%
\providecommand \url  [0]{\begingroup\@sanitize \@url }%
\providecommand \@url [1]{\endgroup\@href {#1}{\urlprefix}}%
\providecommand \urlprefix [0]{URL }%
\providecommand \Eprint[0]{\href }%
\@ifxundefined \urlstyle {%
  \providecommand \doi [1]{doi:\discretionary{}{}{}#1}%
}{%
  \providecommand \doi [0]{doi:\discretionary{}{}{}\begingroup
  \urlstyle{rm}\Url }%
}%
\providecommand \doibase [0]{http://dx.doi.org/}%
\providecommand \Doi[1]{\href{\doibase#1}}%
\providecommand \bibAnnote [3]{%
  \BibitemShut{#1}%
  \begin{quotation}\noindent
    \textsc{Key:}\ #2\\\textsc{Annotation:}\ #3%
  \end{quotation}%
}%
\providecommand \bibAnnoteFile [2]{%
  \IfFileExists{#2}{\bibAnnote {#1} {#2} {\input{#2}}}{}%
}%
\providecommand \typeout [0]{\immediate \write \m@ne }%
\providecommand \selectlanguage [0]{\@gobble}%
\providecommand \bibinfo [0]{\@secondoftwo}%
\providecommand \bibfield [0]{\@secondoftwo}%
\providecommand \translation [1]{[#1]}%
\providecommand \BibitemOpen[0]{}%
\providecommand \bibitemStop [0]{}%
\providecommand \bibitemNoStop [0]{.\EOS\space}%
\providecommand \EOS [0]{\spacefactor3000\relax}%
\providecommand \BibitemShut [1]{\csname bibitem#1\endcsname}%
\bibitem{Komatsu:2008hk}%
  \BibitemOpen
  \bibfield{author}{%
  \bibinfo {author} {\bibfnamefont{E.}~\bibnamefont{Komatsu}} \emph{et~al.}
  (\bibinfo {collaboration} {WMAP}),\ }%
  \bibfield{journal}{%
  \Doi{10.1088/0067-0049/180/2/330}{\bibinfo {journal} {Astrophys. J. Suppl.}}\
  }%
  \textbf{\bibinfo {volume} {180}},\ \bibinfo {pages} {330} (\bibinfo {year}
  {2009}),\ \Eprint{http://arxiv.org/abs/0803.0547}{arXiv:0803.0547
  [astro-ph]}%
  \bibAnnoteFile{NoStop}{Komatsu:2008hk}%
\bibitem{Komatsu:2010fb}%
  \BibitemOpen
  \bibfield{author}{%
  \bibinfo {author} {\bibfnamefont{E.}~\bibnamefont{Komatsu}} \emph{et~al.}}%
   (\bibinfo {year} {2010}),\
  \Eprint{http://arxiv.org/abs/1001.4538}{arXiv:1001.4538 [astro-ph.CO]}%
  \bibAnnoteFile{NoStop}{Komatsu:2010fb}%
\bibitem{Lee:1977ua}%
  \BibitemOpen
  \bibfield{author}{%
  \bibinfo {author} {\bibfnamefont{B.~W.}\ \bibnamefont{Lee}}\ and\ \bibinfo
  {author} {\bibfnamefont{S.}~\bibnamefont{Weinberg}},\ }%
  \bibfield{journal}{%
  \Doi{10.1103/PhysRevLett.39.165}{\bibinfo {journal} {Phys. Rev. Lett.}}\ }%
  \textbf{\bibinfo {volume} {39}},\ \bibinfo {pages} {165} (\bibinfo {year}
  {1977})%
  \bibAnnoteFile{NoStop}{Lee:1977ua}%
\bibitem{Sato:1977ye}%
  \BibitemOpen
  \bibfield{author}{%
  \bibinfo {author} {\bibfnamefont{K.}~\bibnamefont{Sato}}\ and\ \bibinfo
  {author} {\bibfnamefont{M.}~\bibnamefont{Kobayashi}},\ }%
  \bibfield{journal}{%
  \Doi{10.1143/PTP.58.1775}{\bibinfo {journal} {Prog. Theor. Phys.}}\ }%
  \textbf{\bibinfo {volume} {58}},\ \bibinfo {pages} {1775} (\bibinfo {year}
  {1977})%
  \bibAnnoteFile{NoStop}{Sato:1977ye}%
\bibitem{Bertone:2004pz}%
  \BibitemOpen
  \bibfield{author}{%
  \bibinfo {author} {\bibfnamefont{G.}~\bibnamefont{Bertone}}, \bibinfo
  {author} {\bibfnamefont{D.}~\bibnamefont{Hooper}},\ and\ \bibinfo {author}
  {\bibfnamefont{J.}~\bibnamefont{Silk}},\ }%
  \bibfield{journal}{%
  \Doi{10.1016/j.physrep.2004.08.031}{\bibinfo {journal} {Phys. Rept.}}\ }%
  \textbf{\bibinfo {volume} {405}},\ \bibinfo {pages} {279} (\bibinfo {year}
  {2005}),\ \Eprint{http://arxiv.org/abs/hep-ph/0404175}{arXiv:hep-ph/0404175}%
  \bibAnnoteFile{NoStop}{Bertone:2004pz}%
\bibitem{Adriani:2008zr}%
  \BibitemOpen
  \bibfield{author}{%
  \bibinfo {author} {\bibfnamefont{O.}~\bibnamefont{Adriani}} \emph{et~al.}
  (\bibinfo {collaboration} {PAMELA}),\ }%
  \bibfield{journal}{%
  \Doi{10.1038/nature07942}{\bibinfo {journal} {Nature}}\ }%
  \textbf{\bibinfo {volume} {458}},\ \bibinfo {pages} {607} (\bibinfo {year}
  {2009}),\ \Eprint{http://arxiv.org/abs/0810.4995}{arXiv:0810.4995
  [astro-ph]}%
  \bibAnnoteFile{NoStop}{Adriani:2008zr}%
\bibitem{Chang:2008zzr}%
  \BibitemOpen
  \bibfield{author}{%
  \bibinfo {author} {\bibfnamefont{J.}~\bibnamefont{Chang}} \emph{et~al.},\ }%
  \bibfield{journal}{%
  \Doi{10.1038/nature07477}{\bibinfo {journal} {Nature}}\ }%
  \textbf{\bibinfo {volume} {456}},\ \bibinfo {pages} {362} (\bibinfo {year}
  {2008})%
  \bibAnnoteFile{NoStop}{Chang:2008zzr}%
\bibitem{Abdo:2009zk}%
  \BibitemOpen
  \bibfield{author}{%
  \bibinfo {author} {\bibfnamefont{A.~A.}\ \bibnamefont{Abdo}} \emph{et~al.}
  (\bibinfo {collaboration} {The Fermi LAT}),\ }%
  \bibfield{journal}{%
  \Doi{10.1103/PhysRevLett.102.181101}{\bibinfo {journal} {Phys. Rev. Lett.}}\
  }%
  \textbf{\bibinfo {volume} {102}},\ \bibinfo {pages} {181101} (\bibinfo {year}
  {2009}),\ \Eprint{http://arxiv.org/abs/0905.0025}{arXiv:0905.0025
  [astro-ph.HE]}%
  \bibAnnoteFile{NoStop}{Abdo:2009zk}%
\bibitem{Ackermann:2010ij}%
  \BibitemOpen
  \bibfield{author}{%
  \bibinfo {author} {\bibfnamefont{M.}~\bibnamefont{Ackermann}} \emph{et~al.}
  (\bibinfo {collaboration} {Fermi LAT})}%
   (\bibinfo {year} {2010}),\
  \Eprint{http://arxiv.org/abs/1008.3999}{arXiv:1008.3999 [astro-ph.HE]}%
  \bibAnnoteFile{NoStop}{Ackermann:2010ij}%
\bibitem{Profumo:2008ms}%
  \BibitemOpen
  \bibfield{author}{%
  \bibinfo {author} {\bibfnamefont{S.}~\bibnamefont{Profumo}}}%
   (\bibinfo {year} {2008}),\
  \Eprint{http://arxiv.org/abs/0812.4457}{arXiv:0812.4457 [astro-ph]}%
  \bibAnnoteFile{NoStop}{Profumo:2008ms}%
\bibitem{Grasso:2009ma}%
  \BibitemOpen
  \bibfield{author}{%
  \bibinfo {author} {\bibfnamefont{D.}~\bibnamefont{Grasso}} \emph{et~al.}
  (\bibinfo {collaboration} {FERMI-LAT}),\ }%
  \bibfield{journal}{%
  \Doi{10.1016/j.astropartphys.2009.07.003}{\bibinfo {journal} {Astropart.
  Phys.}}\ }%
  \textbf{\bibinfo {volume} {32}},\ \bibinfo {pages} {140} (\bibinfo {year}
  {2009}),\ \Eprint{http://arxiv.org/abs/0905.0636}{arXiv:0905.0636
  [astro-ph.HE]}%
  \bibAnnoteFile{NoStop}{Grasso:2009ma}%
\bibitem{Biermann:2009qi}%
  \BibitemOpen
  \bibfield{author}{%
  \bibinfo {author} {\bibfnamefont{P.~L.}\ \bibnamefont{Biermann}}
  \emph{et~al.},\ }%
  \bibfield{journal}{%
  \Doi{10.1103/PhysRevLett.103.061101}{\bibinfo {journal} {Phys. Rev. Lett.}}\
  }%
  \textbf{\bibinfo {volume} {103}},\ \bibinfo {pages} {061101} (\bibinfo {year}
  {2009}),\ \Eprint{http://arxiv.org/abs/0903.4048}{arXiv:0903.4048
  [astro-ph.HE]}%
  \bibAnnoteFile{NoStop}{Biermann:2009qi}%
\bibitem{Ando:2005xg}%
  \BibitemOpen
  \bibfield{author}{%
  \bibinfo {author} {\bibfnamefont{S.}~\bibnamefont{Ando}}\ and\ \bibinfo
  {author} {\bibfnamefont{E.}~\bibnamefont{Komatsu}},\ }%
  \bibfield{journal}{%
  \Doi{10.1103/PhysRevD.73.023521}{\bibinfo {journal} {Phys. Rev.}}\ }%
  \textbf{\bibinfo {volume} {D73}},\ \bibinfo {pages} {023521} (\bibinfo {year}
  {2006}),\
  \Eprint{http://arxiv.org/abs/astro-ph/0512217}{arXiv:astro-ph/0512217}%
  \bibAnnoteFile{NoStop}{Ando:2005xg}%
\bibitem{Ando:2006cr}%
  \BibitemOpen
  \bibfield{author}{%
  \bibinfo {author} {\bibfnamefont{S.}~\bibnamefont{Ando}}, \bibinfo {author}
  {\bibfnamefont{E.}~\bibnamefont{Komatsu}}, \bibinfo {author}
  {\bibfnamefont{T.}~\bibnamefont{Narumoto}},\ and\ \bibinfo {author}
  {\bibfnamefont{T.}~\bibnamefont{Totani}},\ }%
  \bibfield{journal}{%
  \Doi{10.1103/PhysRevD.75.063519}{\bibinfo {journal} {Phys. Rev.}}\ }%
  \textbf{\bibinfo {volume} {D75}},\ \bibinfo {pages} {063519} (\bibinfo {year}
  {2007}),\
  \Eprint{http://arxiv.org/abs/astro-ph/0612467}{arXiv:astro-ph/0612467}%
  \bibAnnoteFile{NoStop}{Ando:2006cr}%
\bibitem{Cuoco:2007sh}%
  \BibitemOpen
  \bibfield{author}{%
  \bibinfo {author} {\bibfnamefont{A.}~\bibnamefont{Cuoco}}, \bibinfo {author}
  {\bibfnamefont{J.}~\bibnamefont{Brandbyge}}, \bibinfo {author}
  {\bibfnamefont{S.}~\bibnamefont{Hannestad}}, \bibinfo {author}
  {\bibfnamefont{T.}~\bibnamefont{Haugboelle}},\ and\ \bibinfo {author}
  {\bibfnamefont{G.}~\bibnamefont{Miele}}}%
   (\bibinfo {year} {2007}),\
  \Eprint{http://arxiv.org/abs/0710.4136}{arXiv:0710.4136 [astro-ph]}%
  \bibAnnoteFile{NoStop}{Cuoco:2007sh}%
\bibitem{SiegalGaskins:2008ge}%
  \BibitemOpen
  \bibfield{author}{%
  \bibinfo {author} {\bibfnamefont{J.~M.}\ \bibnamefont{Siegal-Gaskins}},\ }%
  \bibfield{journal}{%
  \Doi{10.1088/1475-7516/2008/10/040}{\bibinfo {journal} {JCAP}}\ }%
  \textbf{\bibinfo {volume} {0810}},\ \bibinfo {pages} {040} (\bibinfo {year}
  {2008}),\ \Eprint{http://arxiv.org/abs/0807.1328}{arXiv:0807.1328
  [astro-ph]}%
  \bibAnnoteFile{NoStop}{SiegalGaskins:2008ge}%
\bibitem{Kistler:2009xf}%
  \BibitemOpen
  \bibfield{author}{%
  \bibinfo {author} {\bibfnamefont{M.~D.}\ \bibnamefont{Kistler}}\ and\
  \bibinfo {author} {\bibfnamefont{J.~M.}\ \bibnamefont{Siegal-Gaskins}},\ }%
  \bibfield{journal}{%
  \Doi{10.1103/PhysRevD.81.103521}{\bibinfo {journal} {Phys. Rev.}}\ }%
  \textbf{\bibinfo {volume} {D81}},\ \bibinfo {pages} {103521} (\bibinfo {year}
  {2010}),\ \Eprint{http://arxiv.org/abs/0909.0519}{arXiv:0909.0519
  [astro-ph.HE]}%
  \bibAnnoteFile{NoStop}{Kistler:2009xf}%
\bibitem{Gondolo:1990dk}%
  \BibitemOpen
  \bibfield{author}{%
  \bibinfo {author} {\bibfnamefont{P.}~\bibnamefont{Gondolo}}\ and\ \bibinfo
  {author} {\bibfnamefont{G.}~\bibnamefont{Gelmini}},\ }%
  \bibfield{journal}{%
  \Doi{10.1016/0550-3213(91)90438-4}{\bibinfo {journal} {Nucl. Phys.}}\ }%
  \textbf{\bibinfo {volume} {B360}},\ \bibinfo {pages} {145} (\bibinfo {year}
  {1991})%
  \bibAnnoteFile{NoStop}{Gondolo:1990dk}%
\bibitem{Cooray:2002dia}%
  \BibitemOpen
  \bibfield{author}{%
  \bibinfo {author} {\bibfnamefont{A.}~\bibnamefont{Cooray}}\ and\ \bibinfo
  {author} {\bibfnamefont{R.~K.}\ \bibnamefont{Sheth}},\ }%
  \bibfield{journal}{%
  \Doi{10.1016/S0370-1573(02)00276-4}{\bibinfo {journal} {Phys. Rept.}}\ }%
  \textbf{\bibinfo {volume} {372}},\ \bibinfo {pages} {1} (\bibinfo {year}
  {2002}),\
  \Eprint{http://arxiv.org/abs/astro-ph/0206508}{arXiv:astro-ph/0206508}%
  \bibAnnoteFile{NoStop}{Cooray:2002dia}%
\bibitem{Sheth:1999su}%
  \BibitemOpen
  \bibfield{author}{%
  \bibinfo {author} {\bibfnamefont{R.~K.}\ \bibnamefont{Sheth}}, \bibinfo
  {author} {\bibfnamefont{H.~J.}\ \bibnamefont{Mo}},\ and\ \bibinfo {author}
  {\bibfnamefont{G.}~\bibnamefont{Tormen}},\ }%
  \bibfield{journal}{%
  \Doi{10.1046/j.1365-8711.2001.04006.x}{\bibinfo {journal} {Mon. Not. Roy.
  Astron. Soc.}}\ }%
  \textbf{\bibinfo {volume} {323}},\ \bibinfo {pages} {1} (\bibinfo {year}
  {2001}),\
  \Eprint{http://arxiv.org/abs/astro-ph/9907024}{arXiv:astro-ph/9907024}%
  \bibAnnoteFile{NoStop}{Sheth:1999su}%
\bibitem{Sheth:2001dp}%
  \BibitemOpen
  \bibfield{author}{%
  \bibinfo {author} {\bibfnamefont{R.~K.}\ \bibnamefont{Sheth}}\ and\ \bibinfo
  {author} {\bibfnamefont{G.}~\bibnamefont{Tormen}},\ }%
  \bibfield{journal}{%
  \Doi{10.1046/j.1365-8711.2002.04950.x}{\bibinfo {journal} {Mon. Not. Roy.
  Astron. Soc.}}\ }%
  \textbf{\bibinfo {volume} {329}},\ \bibinfo {pages} {61} (\bibinfo {year}
  {2002}),\
  \Eprint{http://arxiv.org/abs/astro-ph/0105113}{arXiv:astro-ph/0105113}%
  \bibAnnoteFile{NoStop}{Sheth:2001dp}%
\bibitem{White:2002at}%
  \BibitemOpen
  \bibfield{author}{%
  \bibinfo {author} {\bibfnamefont{M.~J.}\ \bibnamefont{White}},\ }%
  \bibfield{journal}{%
  \Doi{10.1086/342752}{\bibinfo {journal} {Astrophys. J. Suppl.}}\ }%
  \textbf{\bibinfo {volume} {143}},\ \bibinfo {pages} {241} (\bibinfo {year}
  {2002}),\
  \Eprint{http://arxiv.org/abs/astro-ph/0207185}{arXiv:astro-ph/0207185}%
  \bibAnnoteFile{NoStop}{White:2002at}%
\bibitem{Navarro:1996gj}%
  \BibitemOpen
  \bibfield{author}{%
  \bibinfo {author} {\bibfnamefont{J.~F.}\ \bibnamefont{Navarro}}, \bibinfo
  {author} {\bibfnamefont{C.~S.}\ \bibnamefont{Frenk}},\ and\ \bibinfo {author}
  {\bibfnamefont{S.~D.~M.}\ \bibnamefont{White}},\ }%
  \bibfield{journal}{%
  \Doi{10.1086/304888}{\bibinfo {journal} {Astrophys. J.}}\ }%
  \textbf{\bibinfo {volume} {490}},\ \bibinfo {pages} {493} (\bibinfo {year}
  {1997}),\
  \Eprint{http://arxiv.org/abs/astro-ph/9611107}{arXiv:astro-ph/9611107}%
  \bibAnnoteFile{NoStop}{Navarro:1996gj}%
\bibitem{Navarro:2008kc}%
  \BibitemOpen
  \bibfield{author}{%
  \bibinfo {author} {\bibfnamefont{J.~F.}\ \bibnamefont{Navarro}}
  \emph{et~al.}}%
   (\bibinfo {year} {2008}),\
  \Eprint{http://arxiv.org/abs/0810.1522}{arXiv:0810.1522 [astro-ph]}%
  \bibAnnoteFile{NoStop}{Navarro:2008kc}%
\bibitem{Duffy:2008pz}%
  \BibitemOpen
  \bibfield{author}{%
  \bibinfo {author} {\bibfnamefont{A.~R.}\ \bibnamefont{Duffy}}, \bibinfo
  {author} {\bibfnamefont{J.}~\bibnamefont{Schaye}}, \bibinfo {author}
  {\bibfnamefont{S.~T.}\ \bibnamefont{Kay}},\ and\ \bibinfo {author}
  {\bibfnamefont{C.}~\bibnamefont{Dalla~Vecchia}},\ }%
  \bibfield{journal}{%
  \bibinfo {journal} {Mon. Not. Roy. Astron. Soc.}\ }%
  \textbf{\bibinfo {volume} {390}},\ \bibinfo {pages} {l64} (\bibinfo {year}
  {2008}),\ \Eprint{http://arxiv.org/abs/0804.2486}{arXiv:0804.2486
  [astro-ph]}%
  \bibAnnoteFile{NoStop}{Duffy:2008pz}%
\bibitem{Bullock:1999he}%
  \BibitemOpen
  \bibfield{author}{%
  \bibinfo {author} {\bibfnamefont{J.~S.}\ \bibnamefont{Bullock}}
  \emph{et~al.},\ }%
  \bibfield{journal}{%
  \Doi{10.1046/j.1365-8711.2001.04068.x}{\bibinfo {journal} {Mon. Not. Roy.
  Astron. Soc.}}\ }%
  \textbf{\bibinfo {volume} {321}},\ \bibinfo {pages} {559} (\bibinfo {year}
  {2001}),\
  \Eprint{http://arxiv.org/abs/astro-ph/9908159}{arXiv:astro-ph/9908159}%
  \bibAnnoteFile{NoStop}{Bullock:1999he}%
\bibitem{Taylor:2001bq}%
  \BibitemOpen
  \bibfield{author}{%
  \bibinfo {author} {\bibfnamefont{J.~E.}\ \bibnamefont{Taylor}}\ and\ \bibinfo
  {author} {\bibfnamefont{J.~F.}\ \bibnamefont{Navarro}},\ }%
  \bibfield{journal}{%
  \Doi{10.1086/324031}{\bibinfo {journal} {Astrophys. J.}}\ }%
  \textbf{\bibinfo {volume} {563}},\ \bibinfo {pages} {483} (\bibinfo {year}
  {2001}),\
  \Eprint{http://arxiv.org/abs/astro-ph/0104002}{arXiv:astro-ph/0104002}%
  \bibAnnoteFile{NoStop}{Taylor:2001bq}%
\bibitem{Stadel:2008pn}%
  \BibitemOpen
  \bibfield{author}{%
  \bibinfo {author} {\bibfnamefont{J.}~\bibnamefont{Stadel}} \emph{et~al.}}%
   (\bibinfo {year} {2008}),\
  \Eprint{http://arxiv.org/abs/0808.2981}{arXiv:0808.2981 [astro-ph]}%
  \bibAnnoteFile{NoStop}{Stadel:2008pn}%
\bibitem{Dehnen:2005cu}%
  \BibitemOpen
  \bibfield{author}{%
  \bibinfo {author} {\bibfnamefont{W.}~\bibnamefont{Dehnen}}\ and\ \bibinfo
  {author} {\bibfnamefont{D.}~\bibnamefont{McLaughlin}},\ }%
  \bibfield{journal}{%
  \Doi{10.1111/j.1365-2966.2005.09510.x}{\bibinfo {journal} {Mon. Not. Roy.
  Astron. Soc.}}\ }%
  \textbf{\bibinfo {volume} {363}},\ \bibinfo {pages} {1057} (\bibinfo {year}
  {2005}),\
  \Eprint{http://arxiv.org/abs/astro-ph/0506528}{arXiv:astro-ph/0506528}%
  \bibAnnoteFile{NoStop}{Dehnen:2005cu}%
\bibitem{Note1}%
  \BibitemOpen
  \bibinfo {note} {More precisely, $v$ is the relative velocity in the center
  of mass frame, but relativistic corrections will be negligible. See \cite
  {Gondolo:1990dk}, for example.}%
  \bibAnnoteFile{Stop}{Note1}%
\bibitem{Drees:1992am}%
  \BibitemOpen
  \bibfield{author}{%
  \bibinfo {author} {\bibfnamefont{M.}~\bibnamefont{Drees}}\ and\ \bibinfo
  {author} {\bibfnamefont{M.~M.}\ \bibnamefont{Nojiri}},\ }%
  \bibfield{journal}{%
  \Doi{10.1103/PhysRevD.47.376}{\bibinfo {journal} {Phys. Rev.}}\ }%
  \textbf{\bibinfo {volume} {D47}},\ \bibinfo {pages} {376} (\bibinfo {year}
  {1993}),\ \Eprint{http://arxiv.org/abs/hep-ph/9207234}{arXiv:hep-ph/9207234}%
  \bibAnnoteFile{NoStop}{Drees:1992am}%
\bibitem{Vogelsberger:2008qb}%
  \BibitemOpen
  \bibfield{author}{%
  \bibinfo {author} {\bibfnamefont{M.}~\bibnamefont{Vogelsberger}}
  \emph{et~al.}}%
   (\bibinfo {year} {2008}),\
  \Eprint{http://arxiv.org/abs/0812.0362}{arXiv:0812.0362 [astro-ph]}%
  \bibAnnoteFile{NoStop}{Vogelsberger:2008qb}%
\bibitem{Zemp:2008gw}%
  \BibitemOpen
  \bibfield{author}{%
  \bibinfo {author} {\bibfnamefont{M.}~\bibnamefont{Zemp}} \emph{et~al.}}%
   (\bibinfo {year} {2008}),\
  \Eprint{http://arxiv.org/abs/0812.2033}{arXiv:0812.2033 [astro-ph]}%
  \bibAnnoteFile{NoStop}{Zemp:2008gw}%
\bibitem{Eisenstein:1997jh}%
  \BibitemOpen
  \bibfield{author}{%
  \bibinfo {author} {\bibfnamefont{D.~J.}\ \bibnamefont{Eisenstein}}\ and\
  \bibinfo {author} {\bibfnamefont{W.}~\bibnamefont{Hu}},\ }%
  \bibfield{journal}{%
  \Doi{10.1086/306640}{\bibinfo {journal} {Astrophys. J.}}\ }%
  \textbf{\bibinfo {volume} {511}},\ \bibinfo {pages} {5} (\bibinfo {year}
  {1997}),\
  \Eprint{http://arxiv.org/abs/astro-ph/9710252}{arXiv:astro-ph/9710252}%
  \bibAnnoteFile{NoStop}{Eisenstein:1997jh}%
\bibitem{Pato:2010yq}%
  \BibitemOpen
  \bibfield{author}{%
  \bibinfo {author} {\bibfnamefont{M.}~\bibnamefont{Pato}}, \bibinfo {author}
  {\bibfnamefont{O.}~\bibnamefont{Agertz}}, \bibinfo {author}
  {\bibfnamefont{G.}~\bibnamefont{Bertone}}, \bibinfo {author}
  {\bibfnamefont{B.}~\bibnamefont{Moore}},\ and\ \bibinfo {author}
  {\bibfnamefont{R.}~\bibnamefont{Teyssier}},\ }%
  \bibfield{journal}{%
  \Doi{10.1103/PhysRevD.82.023531}{\bibinfo {journal} {Phys. Rev.}}\ }%
  \textbf{\bibinfo {volume} {D82}},\ \bibinfo {pages} {023531} (\bibinfo {year}
  {2010}),\ \Eprint{http://arxiv.org/abs/1006.1322}{arXiv:1006.1322
  [astro-ph.HE]}%
  \bibAnnoteFile{NoStop}{Pato:2010yq}%
\bibitem{Bertschinger:2006nq}%
  \BibitemOpen
  \bibfield{author}{%
  \bibinfo {author} {\bibfnamefont{E.}~\bibnamefont{Bertschinger}},\ }%
  \bibfield{journal}{%
  \Doi{10.1103/PhysRevD.74.063509}{\bibinfo {journal} {Phys. Rev.}}\ }%
  \textbf{\bibinfo {volume} {D74}},\ \bibinfo {pages} {063509} (\bibinfo {year}
  {2006}),\
  \Eprint{http://arxiv.org/abs/astro-ph/0607319}{arXiv:astro-ph/0607319}%
  \bibAnnoteFile{NoStop}{Bertschinger:2006nq}%
\bibitem{Bringmann:2009vf}%
  \BibitemOpen
  \bibfield{author}{%
  \bibinfo {author} {\bibfnamefont{T.}~\bibnamefont{Bringmann}},\ }%
  \bibfield{journal}{%
  \Doi{10.1088/1367-2630/11/10/105027}{\bibinfo {journal} {New J. Phys.}}\ }%
  \textbf{\bibinfo {volume} {11}},\ \bibinfo {pages} {105027} (\bibinfo {year}
  {2009}),\ \Eprint{http://arxiv.org/abs/0903.0189}{arXiv:0903.0189
  [astro-ph.CO]}%
  \bibAnnoteFile{NoStop}{Bringmann:2009vf}%
\bibitem{Kasahara:2009th}%
  \BibitemOpen
  \bibfield{author}{%
  \bibinfo {author} {\bibfnamefont{J.}~\bibnamefont{Kasahara}},\ }%
  \emph{\bibinfo {title} {{Neutralino Dark Matter: The Mass of the Smallest
  Halo and the Golden Region}}},\ Ph.D. thesis,\ \bibinfo {school} {University
  of Utah} (\bibinfo {month} {August}\ \bibinfo {year} {2009})%
  \bibAnnoteFile{NoStop}{Kasahara:2009th}%
\bibitem{Diemand:2006ik}%
  \BibitemOpen
  \bibfield{author}{%
  \bibinfo {author} {\bibfnamefont{J.}~\bibnamefont{Diemand}}, \bibinfo
  {author} {\bibfnamefont{M.}~\bibnamefont{Kuhlen}},\ and\ \bibinfo {author}
  {\bibfnamefont{P.}~\bibnamefont{Madau}},\ }%
  \bibfield{journal}{%
  \Doi{10.1086/510736}{\bibinfo {journal} {Astrophys. J.}}\ }%
  \textbf{\bibinfo {volume} {657}},\ \bibinfo {pages} {262} (\bibinfo {year}
  {2007}),\
  \Eprint{http://arxiv.org/abs/astro-ph/0611370}{arXiv:astro-ph/0611370}%
  \bibAnnoteFile{NoStop}{Diemand:2006ik}%
\bibitem{Diemand:2008in}%
  \BibitemOpen
  \bibfield{author}{%
  \bibinfo {author} {\bibfnamefont{J.}~\bibnamefont{Diemand}} \emph{et~al.},\
  }%
  \bibfield{journal}{%
  \Doi{10.1038/nature07153}{\bibinfo {journal} {Nature}}\ }%
  \textbf{\bibinfo {volume} {454}},\ \bibinfo {pages} {735} (\bibinfo {year}
  {2008}),\ \Eprint{http://arxiv.org/abs/0805.1244}{arXiv:0805.1244
  [astro-ph]}%
  \bibAnnoteFile{NoStop}{Diemand:2008in}%
\bibitem{Springel:2008cc}%
  \BibitemOpen
  \bibfield{author}{%
  \bibinfo {author} {\bibfnamefont{V.}~\bibnamefont{Springel}} \emph{et~al.},\
  }%
  \bibfield{journal}{%
  \Doi{10.1111/j.1365-2966.2008.14066.x}{\bibinfo {journal} {Mon. Not. Roy.
  Astron. Soc.}}\ }%
  \textbf{\bibinfo {volume} {391}},\ \bibinfo {pages} {1685} (\bibinfo {year}
  {2008}),\ \Eprint{http://arxiv.org/abs/0809.0898}{arXiv:0809.0898
  [astro-ph]}%
  \bibAnnoteFile{NoStop}{Springel:2008cc}%
\bibitem{Martinez:2009jh}%
  \BibitemOpen
  \bibfield{author}{%
  \bibinfo {author} {\bibfnamefont{G.~D.}\ \bibnamefont{Martinez}}, \bibinfo
  {author} {\bibfnamefont{J.~S.}\ \bibnamefont{Bullock}}, \bibinfo {author}
  {\bibfnamefont{M.}~\bibnamefont{Kaplinghat}}, \bibinfo {author}
  {\bibfnamefont{L.~E.}\ \bibnamefont{Strigari}},\ and\ \bibinfo {author}
  {\bibfnamefont{R.}~\bibnamefont{Trotta}},\ }%
  \bibfield{journal}{%
  \Doi{10.1088/1475-7516/2009/06/014}{\bibinfo {journal} {JCAP}}\ }%
  \textbf{\bibinfo {volume} {0906}},\ \bibinfo {pages} {014} (\bibinfo {year}
  {2009}),\ \Eprint{http://arxiv.org/abs/0902.4715}{arXiv:0902.4715
  [astro-ph.HE]}%
  \bibAnnoteFile{NoStop}{Martinez:2009jh}%
\bibitem{Afshordi:2009hn}%
  \BibitemOpen
  \bibfield{author}{%
  \bibinfo {author} {\bibfnamefont{N.}~\bibnamefont{Afshordi}}, \bibinfo
  {author} {\bibfnamefont{R.}~\bibnamefont{Mohayaee}},\ and\ \bibinfo {author}
  {\bibfnamefont{E.}~\bibnamefont{Bertschinger}}}%
   (\bibinfo {year} {2009}),\
  \Eprint{http://arxiv.org/abs/0911.0414}{arXiv:0911.0414 [astro-ph.CO]}%
  \bibAnnoteFile{NoStop}{Afshordi:2009hn}%
\bibitem{Stecker:2005qs}%
  \BibitemOpen
  \bibfield{author}{%
  \bibinfo {author} {\bibfnamefont{F.~W.}\ \bibnamefont{Stecker}}, \bibinfo
  {author} {\bibfnamefont{M.~A.}\ \bibnamefont{Malkan}},\ and\ \bibinfo
  {author} {\bibfnamefont{S.~T.}\ \bibnamefont{Scully}},\ }%
  \bibfield{journal}{%
  \Doi{10.1086/506188}{\bibinfo {journal} {Astrophys. J.}}\ }%
  \textbf{\bibinfo {volume} {648}},\ \bibinfo {pages} {774} (\bibinfo {year}
  {2006}),\
  \Eprint{http://arxiv.org/abs/astro-ph/0510449}{arXiv:astro-ph/0510449}%
  \bibAnnoteFile{NoStop}{Stecker:2005qs}%
\bibitem{Stecker:2006eh}%
  \BibitemOpen
  \bibfield{author}{%
  \bibinfo {author} {\bibfnamefont{F.~W.}\ \bibnamefont{Stecker}}, \bibinfo
  {author} {\bibfnamefont{M.~A.}\ \bibnamefont{Malkan}},\ and\ \bibinfo
  {author} {\bibfnamefont{S.~T.}\ \bibnamefont{Scully}},\ }%
  \bibfield{journal}{%
  \Doi{10.1086/511738}{\bibinfo {journal} {Astrophys. J.}}\ }%
  \textbf{\bibinfo {volume} {658}},\ \bibinfo {pages} {1392} (\bibinfo {year}
  {2007}),\
  \Eprint{http://arxiv.org/abs/astro-ph/0612048}{arXiv:astro-ph/0612048}%
  \bibAnnoteFile{NoStop}{Stecker:2006eh}%
\bibitem{Note2}%
  \BibitemOpen
  \bibinfo {note} {We do not use the angle bracket notation with the mean
  one-point velocity-weighted cross section to prevent confusion, since that
  notation is traditionally used to denote a thermal average in the literature.
  Also, the angle brackets will predominantly be used in this paper to denote
  averages over ensembles of dark matter halos.}%
  \bibAnnoteFile{Stop}{Note2}%
\bibitem{Note3}%
  \BibitemOpen
  \bibinfo {note} {In the Bullock et al. distribution of halo concentrations,
  there is a halo mass above which the mean halo concentration becomes
  negative.}%
  \bibAnnoteFile{Stop}{Note3}%
\bibitem{Gondolo:2004sc}%
  \BibitemOpen
  \bibfield{author}{%
  \bibinfo {author} {\bibfnamefont{P.}~\bibnamefont{Gondolo}} \emph{et~al.},\
  }%
  \bibfield{journal}{%
  \Doi{10.1088/1475-7516/2004/07/008}{\bibinfo {journal} {JCAP}}\ }%
  \textbf{\bibinfo {volume} {0407}},\ \bibinfo {pages} {008} (\bibinfo {year}
  {2004}),\
  \Eprint{http://arxiv.org/abs/astro-ph/0406204}{arXiv:astro-ph/0406204}%
  \bibAnnoteFile{NoStop}{Gondolo:2004sc}%
\bibitem{Paige:2003mg}%
  \BibitemOpen
  \bibfield{author}{%
  \bibinfo {author} {\bibfnamefont{F.~E.}\ \bibnamefont{Paige}}, \bibinfo
  {author} {\bibfnamefont{S.~D.}\ \bibnamefont{Protopopescu}}, \bibinfo
  {author} {\bibfnamefont{H.}~\bibnamefont{Baer}},\ and\ \bibinfo {author}
  {\bibfnamefont{X.}~\bibnamefont{Tata}}}%
   (\bibinfo {year} {2003}),\
  \Eprint{http://arxiv.org/abs/hep-ph/0312045}{arXiv:hep-ph/0312045}%
  \bibAnnoteFile{NoStop}{Paige:2003mg}%
\bibitem{Hahn:2009zz}%
  \BibitemOpen
  \bibfield{author}{%
  \bibinfo {author} {\bibfnamefont{T.}~\bibnamefont{Hahn}}, \bibinfo {author}
  {\bibfnamefont{S.}~\bibnamefont{Heinemeyer}}, \bibinfo {author}
  {\bibfnamefont{W.}~\bibnamefont{Hollik}}, \bibinfo {author}
  {\bibfnamefont{H.}~\bibnamefont{Rzehak}},\ and\ \bibinfo {author}
  {\bibfnamefont{G.}~\bibnamefont{Weiglein}},\ }%
  \bibfield{journal}{%
  \Doi{10.1016/j.cpc.2009.02.014}{\bibinfo {journal} {Comput. Phys. Commun.}}\
  }%
  \textbf{\bibinfo {volume} {180}},\ \bibinfo {pages} {1426} (\bibinfo {year}
  {2009})%
  \bibAnnoteFile{NoStop}{Hahn:2009zz}%
\bibitem{Chan:1997bi}%
  \BibitemOpen
  \bibfield{author}{%
  \bibinfo {author} {\bibfnamefont{K.~L.}\ \bibnamefont{Chan}}, \bibinfo
  {author} {\bibfnamefont{U.}~\bibnamefont{Chattopadhyay}},\ and\ \bibinfo
  {author} {\bibfnamefont{P.}~\bibnamefont{Nath}},\ }%
  \bibfield{journal}{%
  \Doi{10.1103/PhysRevD.58.096004}{\bibinfo {journal} {Phys. Rev.}}\ }%
  \textbf{\bibinfo {volume} {D58}},\ \bibinfo {pages} {096004} (\bibinfo {year}
  {1998}),\ \Eprint{http://arxiv.org/abs/hep-ph/9710473}{arXiv:hep-ph/9710473}%
  \bibAnnoteFile{NoStop}{Chan:1997bi}%
\bibitem{Feng:1999mn}%
  \BibitemOpen
  \bibfield{author}{%
  \bibinfo {author} {\bibfnamefont{J.~L.}\ \bibnamefont{Feng}}, \bibinfo
  {author} {\bibfnamefont{K.~T.}\ \bibnamefont{Matchev}},\ and\ \bibinfo
  {author} {\bibfnamefont{T.}~\bibnamefont{Moroi}},\ }%
  \bibfield{journal}{%
  \Doi{10.1103/PhysRevLett.84.2322}{\bibinfo {journal} {Phys. Rev. Lett.}}\ }%
  \textbf{\bibinfo {volume} {84}},\ \bibinfo {pages} {2322} (\bibinfo {year}
  {2000}),\ \Eprint{http://arxiv.org/abs/hep-ph/9908309}{arXiv:hep-ph/9908309}%
  \bibAnnoteFile{NoStop}{Feng:1999mn}%
\bibitem{Feng:1999zg}%
  \BibitemOpen
  \bibfield{author}{%
  \bibinfo {author} {\bibfnamefont{J.~L.}\ \bibnamefont{Feng}}, \bibinfo
  {author} {\bibfnamefont{K.~T.}\ \bibnamefont{Matchev}},\ and\ \bibinfo
  {author} {\bibfnamefont{T.}~\bibnamefont{Moroi}},\ }%
  \bibfield{journal}{%
  \Doi{10.1103/PhysRevD.61.075005}{\bibinfo {journal} {Phys. Rev.}}\ }%
  \textbf{\bibinfo {volume} {D61}},\ \bibinfo {pages} {075005} (\bibinfo {year}
  {2000}),\ \Eprint{http://arxiv.org/abs/hep-ph/9909334}{arXiv:hep-ph/9909334}%
  \bibAnnoteFile{NoStop}{Feng:1999zg}%
\bibitem{Baer:1995nq}%
  \BibitemOpen
  \bibfield{author}{%
  \bibinfo {author} {\bibfnamefont{H.}~\bibnamefont{Baer}}, \bibinfo {author}
  {\bibfnamefont{C.-h.}\ \bibnamefont{Chen}}, \bibinfo {author}
  {\bibfnamefont{F.}~\bibnamefont{Paige}},\ and\ \bibinfo {author}
  {\bibfnamefont{X.}~\bibnamefont{Tata}},\ }%
  \bibfield{journal}{%
  \Doi{10.1103/PhysRevD.52.2746}{\bibinfo {journal} {Phys. Rev.}}\ }%
  \textbf{\bibinfo {volume} {D52}},\ \bibinfo {pages} {2746} (\bibinfo {year}
  {1995}),\ \Eprint{http://arxiv.org/abs/hep-ph/9503271}{arXiv:hep-ph/9503271}%
  \bibAnnoteFile{NoStop}{Baer:1995nq}%
\bibitem{Baer:1995va}%
  \BibitemOpen
  \bibfield{author}{%
  \bibinfo {author} {\bibfnamefont{H.}~\bibnamefont{Baer}}, \bibinfo {author}
  {\bibfnamefont{C.-h.}\ \bibnamefont{Chen}}, \bibinfo {author}
  {\bibfnamefont{F.}~\bibnamefont{Paige}},\ and\ \bibinfo {author}
  {\bibfnamefont{X.}~\bibnamefont{Tata}},\ }%
  \bibfield{journal}{%
  \Doi{10.1103/PhysRevD.53.6241}{\bibinfo {journal} {Phys. Rev.}}\ }%
  \textbf{\bibinfo {volume} {D53}},\ \bibinfo {pages} {6241} (\bibinfo {year}
  {1996}),\ \Eprint{http://arxiv.org/abs/hep-ph/9512383}{arXiv:hep-ph/9512383}%
  \bibAnnoteFile{NoStop}{Baer:1995va}%
\bibitem{Baer:1998sz}%
  \BibitemOpen
  \bibfield{author}{%
  \bibinfo {author} {\bibfnamefont{H.}~\bibnamefont{Baer}}, \bibinfo {author}
  {\bibfnamefont{C.-h.}\ \bibnamefont{Chen}}, \bibinfo {author}
  {\bibfnamefont{M.}~\bibnamefont{Drees}}, \bibinfo {author}
  {\bibfnamefont{F.}~\bibnamefont{Paige}},\ and\ \bibinfo {author}
  {\bibfnamefont{X.}~\bibnamefont{Tata}},\ }%
  \bibfield{journal}{%
  \Doi{10.1103/PhysRevD.59.055014}{\bibinfo {journal} {Phys. Rev.}}\ }%
  \textbf{\bibinfo {volume} {D59}},\ \bibinfo {pages} {055014} (\bibinfo {year}
  {1999}),\ \Eprint{http://arxiv.org/abs/hep-ph/9809223}{arXiv:hep-ph/9809223}%
  \bibAnnoteFile{NoStop}{Baer:1998sz}%
\bibitem{Ellis:1998kh}%
  \BibitemOpen
  \bibfield{author}{%
  \bibinfo {author} {\bibfnamefont{J.~R.}\ \bibnamefont{Ellis}}, \bibinfo
  {author} {\bibfnamefont{T.}~\bibnamefont{Falk}},\ and\ \bibinfo {author}
  {\bibfnamefont{K.~A.}\ \bibnamefont{Olive}},\ }%
  \bibfield{journal}{%
  \Doi{10.1016/S0370-2693(98)01392-6}{\bibinfo {journal} {Phys. Lett.}}\ }%
  \textbf{\bibinfo {volume} {B444}},\ \bibinfo {pages} {367} (\bibinfo {year}
  {1998}),\ \Eprint{http://arxiv.org/abs/hep-ph/9810360}{arXiv:hep-ph/9810360}%
  \bibAnnoteFile{NoStop}{Ellis:1998kh}%
\bibitem{Ellis:1999mm}%
  \BibitemOpen
  \bibfield{author}{%
  \bibinfo {author} {\bibfnamefont{J.~R.}\ \bibnamefont{Ellis}}, \bibinfo
  {author} {\bibfnamefont{T.}~\bibnamefont{Falk}}, \bibinfo {author}
  {\bibfnamefont{K.~A.}\ \bibnamefont{Olive}},\ and\ \bibinfo {author}
  {\bibfnamefont{M.}~\bibnamefont{Srednicki}},\ }%
  \bibfield{journal}{%
  \Doi{10.1016/S0927-6505(99)00104-8}{\bibinfo {journal} {Astropart. Phys.}}\
  }%
  \textbf{\bibinfo {volume} {13}},\ \bibinfo {pages} {181} (\bibinfo {year}
  {2000}),\ \Eprint{http://arxiv.org/abs/hep-ph/9905481}{arXiv:hep-ph/9905481}%
  \bibAnnoteFile{NoStop}{Ellis:1999mm}%
\bibitem{Gomez:1999dk}%
  \BibitemOpen
  \bibfield{author}{%
  \bibinfo {author} {\bibfnamefont{M.~E.}\ \bibnamefont{Gomez}}, \bibinfo
  {author} {\bibfnamefont{G.}~\bibnamefont{Lazarides}},\ and\ \bibinfo {author}
  {\bibfnamefont{C.}~\bibnamefont{Pallis}},\ }%
  \bibfield{journal}{%
  \Doi{10.1103/PhysRevD.61.123512}{\bibinfo {journal} {Phys. Rev.}}\ }%
  \textbf{\bibinfo {volume} {D61}},\ \bibinfo {pages} {123512} (\bibinfo {year}
  {2000}),\ \Eprint{http://arxiv.org/abs/hep-ph/9907261}{arXiv:hep-ph/9907261}%
  \bibAnnoteFile{NoStop}{Gomez:1999dk}%
\bibitem{Gomez:2000sj}%
  \BibitemOpen
  \bibfield{author}{%
  \bibinfo {author} {\bibfnamefont{M.~E.}\ \bibnamefont{Gomez}}, \bibinfo
  {author} {\bibfnamefont{G.}~\bibnamefont{Lazarides}},\ and\ \bibinfo {author}
  {\bibfnamefont{C.}~\bibnamefont{Pallis}},\ }%
  \bibfield{journal}{%
  \Doi{10.1016/S0370-2693(00)00841-8}{\bibinfo {journal} {Phys. Lett.}}\ }%
  \textbf{\bibinfo {volume} {B487}},\ \bibinfo {pages} {313} (\bibinfo {year}
  {2000}),\ \Eprint{http://arxiv.org/abs/hep-ph/0004028}{arXiv:hep-ph/0004028}%
  \bibAnnoteFile{NoStop}{Gomez:2000sj}%
\bibitem{Lahanas:1999uy}%
  \BibitemOpen
  \bibfield{author}{%
  \bibinfo {author} {\bibfnamefont{A.~B.}\ \bibnamefont{Lahanas}}, \bibinfo
  {author} {\bibfnamefont{D.~V.}\ \bibnamefont{Nanopoulos}},\ and\ \bibinfo
  {author} {\bibfnamefont{V.~C.}\ \bibnamefont{Spanos}},\ }%
  \bibfield{journal}{%
  \Doi{10.1103/PhysRevD.62.023515}{\bibinfo {journal} {Phys. Rev.}}\ }%
  \textbf{\bibinfo {volume} {D62}},\ \bibinfo {pages} {023515} (\bibinfo {year}
  {2000}),\ \Eprint{http://arxiv.org/abs/hep-ph/9909497}{arXiv:hep-ph/9909497}%
  \bibAnnoteFile{NoStop}{Lahanas:1999uy}%
\bibitem{Arnowitt:2001yh}%
  \BibitemOpen
  \bibfield{author}{%
  \bibinfo {author} {\bibfnamefont{R.~L.}\ \bibnamefont{Arnowitt}}, \bibinfo
  {author} {\bibfnamefont{B.}~\bibnamefont{Dutta}},\ and\ \bibinfo {author}
  {\bibfnamefont{Y.}~\bibnamefont{Santoso}},\ }%
  \bibfield{journal}{%
  \Doi{10.1016/S0550-3213(01)00230-9}{\bibinfo {journal} {Nucl. Phys.}}\ }%
  \textbf{\bibinfo {volume} {B606}},\ \bibinfo {pages} {59} (\bibinfo {year}
  {2001}),\ \Eprint{http://arxiv.org/abs/hep-ph/0102181}{arXiv:hep-ph/0102181}%
  \bibAnnoteFile{NoStop}{Arnowitt:2001yh}%
\bibitem{Baer:1997ai}%
  \BibitemOpen
  \bibfield{author}{%
  \bibinfo {author} {\bibfnamefont{H.}~\bibnamefont{Baer}}\ and\ \bibinfo
  {author} {\bibfnamefont{M.}~\bibnamefont{Brhlik}},\ }%
  \bibfield{journal}{%
  \Doi{10.1103/PhysRevD.57.567}{\bibinfo {journal} {Phys. Rev.}}\ }%
  \textbf{\bibinfo {volume} {D57}},\ \bibinfo {pages} {567} (\bibinfo {year}
  {1998}),\ \Eprint{http://arxiv.org/abs/hep-ph/9706509}{arXiv:hep-ph/9706509}%
  \bibAnnoteFile{NoStop}{Baer:1997ai}%
\bibitem{Baer:2000jj}%
  \BibitemOpen
  \bibfield{author}{%
  \bibinfo {author} {\bibfnamefont{H.}~\bibnamefont{Baer}} \emph{et~al.},\ }%
  \bibfield{journal}{%
  \Doi{10.1103/PhysRevD.63.015007}{\bibinfo {journal} {Phys. Rev.}}\ }%
  \textbf{\bibinfo {volume} {D63}},\ \bibinfo {pages} {015007} (\bibinfo {year}
  {2001}),\ \Eprint{http://arxiv.org/abs/hep-ph/0005027}{arXiv:hep-ph/0005027}%
  \bibAnnoteFile{NoStop}{Baer:2000jj}%
\bibitem{Ellis:2001msa}%
  \BibitemOpen
  \bibfield{author}{%
  \bibinfo {author} {\bibfnamefont{J.~R.}\ \bibnamefont{Ellis}}, \bibinfo
  {author} {\bibfnamefont{T.}~\bibnamefont{Falk}}, \bibinfo {author}
  {\bibfnamefont{G.}~\bibnamefont{Ganis}}, \bibinfo {author}
  {\bibfnamefont{K.~A.}\ \bibnamefont{Olive}},\ and\ \bibinfo {author}
  {\bibfnamefont{M.}~\bibnamefont{Srednicki}},\ }%
  \bibfield{journal}{%
  \Doi{10.1016/S0370-2693(01)00541-X}{\bibinfo {journal} {Phys. Lett.}}\ }%
  \textbf{\bibinfo {volume} {B510}},\ \bibinfo {pages} {236} (\bibinfo {year}
  {2001}),\ \Eprint{http://arxiv.org/abs/hep-ph/0102098}{arXiv:hep-ph/0102098}%
  \bibAnnoteFile{NoStop}{Ellis:2001msa}%
\bibitem{Roszkowski:2001sb}%
  \BibitemOpen
  \bibfield{author}{%
  \bibinfo {author} {\bibfnamefont{L.}~\bibnamefont{Roszkowski}}, \bibinfo
  {author} {\bibfnamefont{R.}~\bibnamefont{Ruiz~de Austri}},\ and\ \bibinfo
  {author} {\bibfnamefont{T.}~\bibnamefont{Nihei}},\ }%
  \bibfield{journal}{%
  \bibinfo {journal} {JHEP}\ }%
  \textbf{\bibinfo {volume} {08}},\ \bibinfo {pages} {024} (\bibinfo {year}
  {2001}),\ \Eprint{http://arxiv.org/abs/hep-ph/0106334}{arXiv:hep-ph/0106334}%
  \bibAnnoteFile{NoStop}{Roszkowski:2001sb}%
\bibitem{Djouadi:2001yk}%
  \BibitemOpen
  \bibfield{author}{%
  \bibinfo {author} {\bibfnamefont{A.}~\bibnamefont{Djouadi}}, \bibinfo
  {author} {\bibfnamefont{M.}~\bibnamefont{Drees}},\ and\ \bibinfo {author}
  {\bibfnamefont{J.~L.}\ \bibnamefont{Kneur}},\ }%
  \bibfield{journal}{%
  \bibinfo {journal} {JHEP}\ }%
  \textbf{\bibinfo {volume} {08}},\ \bibinfo {pages} {055} (\bibinfo {year}
  {2001}),\ \Eprint{http://arxiv.org/abs/hep-ph/0107316}{arXiv:hep-ph/0107316}%
  \bibAnnoteFile{NoStop}{Djouadi:2001yk}%
\bibitem{Lahanas:2001yr}%
  \BibitemOpen
  \bibfield{author}{%
  \bibinfo {author} {\bibfnamefont{A.~B.}\ \bibnamefont{Lahanas}}\ and\
  \bibinfo {author} {\bibfnamefont{V.~C.}\ \bibnamefont{Spanos}},\ }%
  \bibfield{journal}{%
  \Doi{10.1007/s100520100861}{\bibinfo {journal} {Eur. Phys. J.}}\ }%
  \textbf{\bibinfo {volume} {C23}},\ \bibinfo {pages} {185} (\bibinfo {year}
  {2002}),\ \Eprint{http://arxiv.org/abs/hep-ph/0106345}{arXiv:hep-ph/0106345}%
  \bibAnnoteFile{NoStop}{Lahanas:2001yr}%
\bibitem{Dutta:2009uf}%
  \BibitemOpen
  \bibfield{author}{%
  \bibinfo {author} {\bibfnamefont{B.}~\bibnamefont{Dutta}}, \bibinfo {author}
  {\bibfnamefont{L.}~\bibnamefont{Leblond}},\ and\ \bibinfo {author}
  {\bibfnamefont{K.}~\bibnamefont{Sinha}},\ }%
  \bibfield{journal}{%
  \Doi{10.1103/PhysRevD.80.035014}{\bibinfo {journal} {Phys. Rev.}}\ }%
  \textbf{\bibinfo {volume} {D80}},\ \bibinfo {pages} {035014} (\bibinfo {year}
  {2009}),\ \Eprint{http://arxiv.org/abs/0904.3773}{arXiv:0904.3773 [hep-ph]}%
  \bibAnnoteFile{NoStop}{Dutta:2009uf}%
\bibitem{Bergstrom:1997fh}%
  \BibitemOpen
  \bibfield{author}{%
  \bibinfo {author} {\bibfnamefont{L.}~\bibnamefont{Bergstrom}}\ and\ \bibinfo
  {author} {\bibfnamefont{P.}~\bibnamefont{Ullio}},\ }%
  \bibfield{journal}{%
  \Doi{10.1016/S0550-3213(97)00530-0}{\bibinfo {journal} {Nucl. Phys.}}\ }%
  \textbf{\bibinfo {volume} {B504}},\ \bibinfo {pages} {27} (\bibinfo {year}
  {1997}),\ \Eprint{http://arxiv.org/abs/hep-ph/9706232}{arXiv:hep-ph/9706232}%
  \bibAnnoteFile{NoStop}{Bergstrom:1997fh}%
\bibitem{Mohapatra:1980qe}%
  \BibitemOpen
  \bibfield{author}{%
  \bibinfo {author} {\bibfnamefont{R.~N.}\ \bibnamefont{Mohapatra}}\ and\
  \bibinfo {author} {\bibfnamefont{R.~E.}\ \bibnamefont{Marshak}},\ }%
  \bibfield{journal}{%
  \Doi{10.1103/PhysRevLett.44.1316}{\bibinfo {journal} {Phys. Rev. Lett.}}\ }%
  \textbf{\bibinfo {volume} {44}},\ \bibinfo {pages} {1316} (\bibinfo {year}
  {1980})%
  \bibAnnoteFile{NoStop}{Mohapatra:1980qe}%
\bibitem{Mohapatra:1980:qf}%
  \BibitemOpen
  \bibfield{author}{%
  \bibinfo {author} {\bibfnamefont{R.~N.}\ \bibnamefont{Mohapatra}}\ and\
  \bibinfo {author} {\bibfnamefont{R.~E.}\ \bibnamefont{Marshak}},\ }%
  \bibfield{journal}{%
  \bibinfo {journal} {Phys. Rev. Lett.}\ }%
  \textbf{\bibinfo {volume} {44}},\ \bibinfo {pages} {1643} (\bibinfo {year}
  {1980})%
  \bibAnnoteFile{NoStop}{Mohapatra:1980:qf}%
\bibitem{Sjostrand:2007gs}%
  \BibitemOpen
  \bibfield{author}{%
  \bibinfo {author} {\bibfnamefont{T.}~\bibnamefont{Sjostrand}}, \bibinfo
  {author} {\bibfnamefont{S.}~\bibnamefont{Mrenna}},\ and\ \bibinfo {author}
  {\bibfnamefont{P.~Z.}\ \bibnamefont{Skands}},\ }%
  \bibfield{journal}{%
  \Doi{10.1016/j.cpc.2008.01.036}{\bibinfo {journal} {Comput. Phys. Commun.}}\
  }%
  \textbf{\bibinfo {volume} {178}},\ \bibinfo {pages} {852} (\bibinfo {year}
  {2008}),\ \Eprint{http://arxiv.org/abs/0710.3820}{arXiv:0710.3820 [hep-ph]}%
  \bibAnnoteFile{NoStop}{Sjostrand:2007gs}%
\bibitem{Allahverdi:2009ae}%
  \BibitemOpen
  \bibfield{author}{%
  \bibinfo {author} {\bibfnamefont{R.}~\bibnamefont{Allahverdi}}, \bibinfo
  {author} {\bibfnamefont{B.}~\bibnamefont{Dutta}}, \bibinfo {author}
  {\bibfnamefont{K.}~\bibnamefont{Richardson-McDaniel}},\ and\ \bibinfo
  {author} {\bibfnamefont{Y.}~\bibnamefont{Santoso}},\ }%
  \bibfield{journal}{%
  \Doi{10.1016/j.physletb.2009.05.034}{\bibinfo {journal} {Phys. Lett.}}\ }%
  \textbf{\bibinfo {volume} {B677}},\ \bibinfo {pages} {172} (\bibinfo {year}
  {2009}),\ \Eprint{http://arxiv.org/abs/0902.3463}{arXiv:0902.3463 [hep-ph]}%
  \bibAnnoteFile{NoStop}{Allahverdi:2009ae}%
\bibitem{Note4}%
  \BibitemOpen
  \bibinfo {note} {This expression is planned to be justified in an upcoming
  paper}%
  \bibAnnoteFile{NoStop}{Note4}%
\bibitem{Hisano:2004ds}%
  \BibitemOpen
  \bibfield{author}{%
  \bibinfo {author} {\bibfnamefont{J.}~\bibnamefont{Hisano}}, \bibinfo {author}
  {\bibfnamefont{S.}~\bibnamefont{Matsumoto}}, \bibinfo {author}
  {\bibfnamefont{M.~M.}\ \bibnamefont{Nojiri}},\ and\ \bibinfo {author}
  {\bibfnamefont{O.}~\bibnamefont{Saito}},\ }%
  \bibfield{journal}{%
  \Doi{10.1103/PhysRevD.71.063528}{\bibinfo {journal} {Phys. Rev.}}\ }%
  \textbf{\bibinfo {volume} {D71}},\ \bibinfo {pages} {063528} (\bibinfo {year}
  {2005}),\ \Eprint{http://arxiv.org/abs/hep-ph/0412403}{arXiv:hep-ph/0412403}%
  \bibAnnoteFile{NoStop}{Hisano:2004ds}%
\bibitem{ArkaniHamed:2008qn}%
  \BibitemOpen
  \bibfield{author}{%
  \bibinfo {author} {\bibfnamefont{N.}~\bibnamefont{Arkani-Hamed}}, \bibinfo
  {author} {\bibfnamefont{D.~P.}\ \bibnamefont{Finkbeiner}}, \bibinfo {author}
  {\bibfnamefont{T.~R.}\ \bibnamefont{Slatyer}},\ and\ \bibinfo {author}
  {\bibfnamefont{N.}~\bibnamefont{Weiner}},\ }%
  \bibfield{journal}{%
  \Doi{10.1103/PhysRevD.79.015014}{\bibinfo {journal} {Phys. Rev.}}\ }%
  \textbf{\bibinfo {volume} {D79}},\ \bibinfo {pages} {015014} (\bibinfo {year}
  {2009}),\ \Eprint{http://arxiv.org/abs/0810.0713}{arXiv:0810.0713 [hep-ph]}%
  \bibAnnoteFile{NoStop}{ArkaniHamed:2008qn}%
\bibitem{Lattanzi:2008qa}%
  \BibitemOpen
  \bibfield{author}{%
  \bibinfo {author} {\bibfnamefont{M.}~\bibnamefont{Lattanzi}}\ and\ \bibinfo
  {author} {\bibfnamefont{J.~I.}\ \bibnamefont{Silk}},\ }%
  \bibfield{journal}{%
  \Doi{10.1103/PhysRevD.79.083523}{\bibinfo {journal} {Phys. Rev.}}\ }%
  \textbf{\bibinfo {volume} {D79}},\ \bibinfo {pages} {083523} (\bibinfo {year}
  {2009}),\ \Eprint{http://arxiv.org/abs/0812.0360}{arXiv:0812.0360
  [astro-ph]}%
  \bibAnnoteFile{NoStop}{Lattanzi:2008qa}%
\bibitem{MarchRussell:2008tu}%
  \BibitemOpen
  \bibfield{author}{%
  \bibinfo {author} {\bibfnamefont{J.~D.}\ \bibnamefont{March-Russell}}\ and\
  \bibinfo {author} {\bibfnamefont{S.~M.}\ \bibnamefont{West}},\ }%
  \bibfield{journal}{%
  \Doi{10.1016/j.physletb.2009.04.010}{\bibinfo {journal} {Phys. Lett.}}\ }%
  \textbf{\bibinfo {volume} {B676}},\ \bibinfo {pages} {133} (\bibinfo {year}
  {2009}),\ \Eprint{http://arxiv.org/abs/0812.0559}{arXiv:0812.0559
  [astro-ph]}%
  \bibAnnoteFile{NoStop}{MarchRussell:2008tu}%
\bibitem{Iengo:2009ni}%
  \BibitemOpen
  \bibfield{author}{%
  \bibinfo {author} {\bibfnamefont{R.}~\bibnamefont{Iengo}},\ }%
  \bibfield{journal}{%
  \Doi{10.1088/1126-6708/2009/05/024}{\bibinfo {journal} {JHEP}}\ }%
  \textbf{\bibinfo {volume} {05}},\ \bibinfo {pages} {024} (\bibinfo {year}
  {2009}),\ \Eprint{http://arxiv.org/abs/0902.0688}{arXiv:0902.0688 [hep-ph]}%
  \bibAnnoteFile{NoStop}{Iengo:2009ni}%
\bibitem{Iengo:2009xf}%
  \BibitemOpen
  \bibfield{author}{%
  \bibinfo {author} {\bibfnamefont{R.}~\bibnamefont{Iengo}}}%
   (\bibinfo {year} {2009}),\
  \Eprint{http://arxiv.org/abs/0903.0317}{arXiv:0903.0317 [hep-ph]}%
  \bibAnnoteFile{NoStop}{Iengo:2009xf}%
\bibitem{Cassel:2009wt}%
  \BibitemOpen
  \bibfield{author}{%
  \bibinfo {author} {\bibfnamefont{S.}~\bibnamefont{Cassel}}}%
   (\bibinfo {year} {2009}),\
  \Eprint{http://arxiv.org/abs/0903.5307}{arXiv:0903.5307 [hep-ph]}%
  \bibAnnoteFile{NoStop}{Cassel:2009wt}%
\bibitem{Feng:2010zp}%
  \BibitemOpen
  \bibfield{author}{%
  \bibinfo {author} {\bibfnamefont{J.~L.}\ \bibnamefont{Feng}}, \bibinfo
  {author} {\bibfnamefont{M.}~\bibnamefont{Kaplinghat}},\ and\ \bibinfo
  {author} {\bibfnamefont{H.-B.}\ \bibnamefont{Yu}}}%
   (\bibinfo {year} {2010}),\
  \Eprint{http://arxiv.org/abs/1005.4678}{arXiv:1005.4678 [hep-ph]}%
  \bibAnnoteFile{NoStop}{Feng:2010zp}%
\bibitem{Backovic:2009rw}%
  \BibitemOpen
  \bibfield{author}{%
  \bibinfo {author} {\bibfnamefont{M.}~\bibnamefont{Backovic}}\ and\ \bibinfo
  {author} {\bibfnamefont{J.~P.}\ \bibnamefont{Ralston}},\ }%
  \bibfield{journal}{%
  \Doi{10.1103/PhysRevD.81.056002}{\bibinfo {journal} {Phys. Rev.}}\ }%
  \textbf{\bibinfo {volume} {D81}},\ \bibinfo {pages} {056002} (\bibinfo {year}
  {2010}),\ \Eprint{http://arxiv.org/abs/0910.1113}{arXiv:0910.1113 [hep-ph]}%
  \bibAnnoteFile{NoStop}{Backovic:2009rw}%
\bibitem{Hui:2001wy}%
  \BibitemOpen
  \bibfield{author}{%
  \bibinfo {author} {\bibfnamefont{L.}~\bibnamefont{Hui}},\ }%
  \bibfield{journal}{%
  \Doi{10.1103/PhysRevLett.86.3467}{\bibinfo {journal} {Phys. Rev. Lett.}}\ }%
  \textbf{\bibinfo {volume} {86}},\ \bibinfo {pages} {3467} (\bibinfo {year}
  {2001}),\
  \Eprint{http://arxiv.org/abs/astro-ph/0102349}{arXiv:astro-ph/0102349}%
  \bibAnnoteFile{NoStop}{Hui:2001wy}%
\bibitem{Note5}%
  \BibitemOpen
  \bibinfo {note} {Work in progress.}%
  \bibAnnoteFile{Stop}{Note5}%
\bibitem{Ibe:2008ye}%
  \BibitemOpen
  \bibfield{author}{%
  \bibinfo {author} {\bibfnamefont{M.}~\bibnamefont{Ibe}}, \bibinfo {author}
  {\bibfnamefont{H.}~\bibnamefont{Murayama}},\ and\ \bibinfo {author}
  {\bibfnamefont{T.~T.}\ \bibnamefont{Yanagida}},\ }%
  \bibfield{journal}{%
  \Doi{10.1103/PhysRevD.79.095009}{\bibinfo {journal} {Phys. Rev.}}\ }%
  \textbf{\bibinfo {volume} {D79}},\ \bibinfo {pages} {095009} (\bibinfo {year}
  {2009}),\ \Eprint{http://arxiv.org/abs/0812.0072}{arXiv:0812.0072 [hep-ph]}%
  \bibAnnoteFile{NoStop}{Ibe:2008ye}%
\bibitem{Peskin:1995ev}%
  \BibitemOpen
  \bibfield{author}{%
  \bibinfo {author} {\bibfnamefont{M.~E.}\ \bibnamefont{Peskin}}\ and\ \bibinfo
  {author} {\bibfnamefont{D.~V.}\ \bibnamefont{Schroeder}},\ }%
  \enquote{\bibinfo {title} {{An Introduction to Quantum Field Theory}},}\ \
  (\bibinfo {publisher} {Westview Press},\ \bibinfo {year} {1995})\ pp.\
  \bibinfo {pages} {101,237}%
  \bibAnnoteFile{NoStop}{Peskin:1995ev}%
\bibitem{Sakurai:1967}%
  \BibitemOpen
  \bibfield{author}{%
  \bibinfo {author} {\bibfnamefont{J.~J.}\ \bibnamefont{Sakurai}},\ }%
  \enquote{\bibinfo {title} {{Advanced Quantum Mechanics}},}\ \ (\bibinfo
  {publisher} {Addison-Wesley},\ \bibinfo {year} {1994})\ Chap.\ \bibinfo
  {chapter} {7.8},\ \bibinfo {edition} {rev.}\ ed.%
  \bibAnnoteFile{Stop}{Sakurai:1967}%
\end{thebibliography}


%

\end{document}